\documentclass[preprint,showpacs,preprintnumbers,amsmath,amssymb]{revtex4}
\usepackage{epsfig}
\usepackage{graphicx}
\usepackage{dcolumn}
\usepackage{bm}
\usepackage{threeparttable}

\def\beq{\begin{equation}}
\def\eeq{\end{equation}}
\def\eeqn{\end{equation}}
\newcommand\iden{\leavevmode\hbox{\small1\normalsize\kern-.33em1}}

\newcommand{\bea} {\begin{eqnarray}}
\newcommand{\eea} {\end{eqnarray}}

\let\jnfont=\rm
\def\NPB#1 {{\jnfont Nucl.\ Phys.\ B }{\bf #1} }
\def\PLB#1 {{\jnfont Phys.\ Lett.\ B }{\bf #1} }
\def\EPJC#1 {{\jnfont Eur.\ Phys.\ Jour.\ C }{\bf #1} }
\def\PRD#1 {{\jnfont Phys.\ Rev.\ D }{\bf #1} }
\def\PRL#1 {{\jnfont Phys.\ Rev.\ Lett.\ }{\bf #1} }
\def\MPLA#1 {{\jnfont Mod.\ Phys.\ Lett.\ A }{\bf #1} }
\def\JPG#1 {{\jnfont J.\ Phys.\ G }{\bf #1} }
\def\CTP#1 {{\jnfont Commun.\ Theor.\ Phys.\ }{\bf #1} }
\def\JHEP#1 {{\jnfont JHEP \ }{\bf #1} }
\def\NPPS#1 {{\jnfont Nucl.\ Phys.\ Proc.\ Suppl.\ }{\bf #1} }
\def\CPC#1 {{\jnfont Comput.\ Phys.\ Commun.\ }{\bf #1} }
\def\CPL#1 {{\jnfont Chin.\ Phys.\ Lett. }{\bf #1} }
\def\APPB#1 {{\jnfont Acta\ Phys.\ Polon.\ B }{\bf #1} }

\def\lsim{\raise0.3ex\hbox{$<$\kern-0.75em\raise-1.1ex\hbox{$\sim$}}}
\def\gsim{\raise0.3ex\hbox{$>$\kern-0.75em\raise-1.1ex\hbox{$\sim$}}}
\def\PR#1 {{\jnfont Phys.\ Rept. }{\bf #1} }
\def\CHC#1 {{\jnfont Chin.\ Phys.\ C }{\bf #1} }
\def\NIMA#1 {{\jnfont Nucl.\ Instrum.\ Meth.\ A }{\bf #1} }
\def\JCAP#1 {{\jnfont JCAP \ }{\bf #1} }
\def\ASA#1 {{\jnfont Astron.\ Astrophys.\ A }{\bf #1} }  

\begin{document}

\title{\ \\[10mm] A light scalar dark matter extension of the type-II two-Higgs-doublet model}

\author{Xiao-Fang Han, Lei Wang}

\affiliation{ Department of Physics, Yantai University, Yantai
264005, China}


\begin{abstract}
We examine the type-II two-Higgs-doublet model with a light scalar dark matter ($S$) after 
imposing the constraints from the Higgs searches at the LHC
and dark matter experiments. We first assume that both two CP-even Higgses ($h$ and $H$) are portals between
the DM and SM sectors, and the CP-odd Higgs ($A$) and $H$ are heavier than 130 GeV. We find that the DM with a mass of $10\sim 50$ GeV
is disfavored by the joint constraints of the 125 GeV Higgs signal data, the relic density, XENON1T (2017), PandaX-II (2017) and
the Fermi-LAT. Next, we consider a special scenario in which the heavy CP-even Higgs is taken as the 125 GeV Higgs.
The light CP-even Higgs is the only portal between the DM and SM sectors, and the DM mass is slightly
below Higgs resonance. We find that the signal data of the 125 GeV Higgs
restrict $\tan\beta$ to be in the range of $1\sim 1.5$ for $m_h<$ 62 GeV. 
The $gg\to A\to hZ$ and $b\bar{b}\to h \to \tau^+\tau^-$ channels at the LHC can impose
lower limits and upper limits on $\tan\beta$, respectively. For appropriate values of $\tan\beta$, $\lambda_h$ and $m_h$, 
the DM with a mass of $10\sim 50$ GeV is allowed by the constraints of the Higgs searches at the LHC
and dark matter experiments. For example, $\tan\beta$ is restricted to be in
the range of $1.0\sim1.5$ for 10 GeV $<m_s<$ 26 GeV, and $\frac{m_h}{2m_S}>$ 1.125 is excluded for 30 GeV $<m_S<$ 50 GeV.

\end{abstract}

\maketitle

\section{Introduction}
The weakly interacting massive particle (WIMP) is one popular candidate of dark matter (DM).
The simplest WIMP-DM model is the standard model (SM) plus a real singlet scalar as DM \cite{1207.4930-1}.
In the model, the current experiments excluded the DM mass up to 330 GeV, except a small range near 63 GeV
\cite{1412.1105,2hisos-6}. Much of the region excluded in this model can be recovered 
if the Higgs sector is extended to the two-Higgs-doublet model (2HDM) \cite{2hdm} which contains
two neutral CP-even Higgs bosons $h$ and $H$, one neutral pseudoscalar $A$, and two charged Higgs $H^{\pm}$ 
\cite{2hisos-6,2hisos-0,2hisos-1,2hisos-2,2hisos-3,2hisos-4,2hisos-5,dmbu,1708.06882,1608.00421}. 
Recently, Ref. \cite{1708.06882} took the 125 GeV Higgs with wrong sign Yukawa coupling of the down-type
quark as the only portal between the DM and SM sector, and found that the DM mass is allowed to 
be as low as 50 GeV for appropriate isospin-violating DM interactions with nucleons. The $SS\to AA$ annihilation channel
 can play an important contribution to the relic density, but does not solve the tension between 
the DM relic density and the signal data of the 125 GeV Higgs,  which leads that $m_S<50$ GeV is excluded.
Ref. \cite{2hisos-6} showed that if the heavy CP-even Higgs boson is the only portal,
much of the region below 100 GeV are excluded.

In this paper, the question we want to answer is, which parameter space of 
the type-II 2HDM with a scalar DM is the DM with
a mass below 50 GeV allowed in? We will consider joint constraints from the theory, the precision electroweak data,
the flavor observables, the signal data of the 125 GeV Higgs, the searches
for the additional Higgs at the LEP and LHC, the relic density,  
XENON1T (2017), PandaX-II (2017) and the Fermi-LAT searches for DM annihilation from 
dwarf spheroidal satellite galaxies (dSphs).

This paper is organized as follows. In Section II, we introduce some characteristic features
of the type-II 2HDM with a scalar DM. In Section III we perform
numerical calculations. In Section IV, we examine the allowed parameter space
after imposing the relevant theoretical and
experimental constraints. Finally, we draw our conclusion in Section V.

\section{Type-II two-Higgs-doublet model with a scalar dark matter}
\subsection{Type-II two-Higgs-doublet model}
In the type-II 2HDM with a scalar DM, 
the scalar potential includes two parts, $\mathcal{V}_{2HDM} +
\mathcal{V}_{S}$, and they are
the potential of type-II 2HDM and the potential of the DM sector, respectively.
The $\mathcal{V}_{2HDM}$ with a softly-broken discrete $Z_2$ symmetry is given by \cite{2h-poten}
\begin{eqnarray} \label{V2HDM} \mathcal{V}_{2HDM} &=& m_{11}^2
(\Phi_1^{\dagger} \Phi_1) + m_{22}^2 (\Phi_2^{\dagger}
\Phi_2) - \left[m_{12}^2 (\Phi_1^{\dagger} \Phi_2 + \rm h.c.)\right]\nonumber \\
&&+ \frac{\lambda_1}{2}  (\Phi_1^{\dagger} \Phi_1)^2 +
\frac{\lambda_2}{2} (\Phi_2^{\dagger} \Phi_2)^2 + \lambda_3
(\Phi_1^{\dagger} \Phi_1)(\Phi_2^{\dagger} \Phi_2) + \lambda_4
(\Phi_1^{\dagger}
\Phi_2)(\Phi_2^{\dagger} \Phi_1) \nonumber \\
&&+ \left[\frac{\lambda_5}{2} (\Phi_1^{\dagger} \Phi_2)^2 + \rm
h.c.\right].
\end{eqnarray}
Here we focus on the case of the CP-conserving in which all $\lambda_i$ and
$m_{12}^2$ are real.
The two complex Higgs doublets have hypercharge $Y = 1$,
\begin{equation}
\Phi_1=\left(\begin{array}{c} \phi_1^+ \\
\frac{1}{\sqrt{2}}\,(v_1+\phi_1^0+ia_1)
\end{array}\right)\,, \ \ \
\Phi_2=\left(\begin{array}{c} \phi_2^+ \\
\frac{1}{\sqrt{2}}\,(v_2+\phi_2^0+ia_2)
\end{array}\right).
\end{equation}
Where $v_1$ and $v_2$ are the electroweak vacuum expectation values
(VEVs) with $v^2 = v^2_1 + v^2_2 = (246~\rm GeV)^2$ and $\tan\beta=v_2 /v_1$. 
After spontaneous electroweak symmetry breaking, the remaining five physical Higgs particles are two neutral
CP-even $h$ and $H$, one neutral pseudoscalar $A$, and two charged
scalars $H^{\pm}$.

In the type II 2HDM, the up-type fermions obtain masses from only $\Phi_2$ field, and the down-type fermions from
$\Phi_1$ field \cite{i-1,ii-2}. The Yukawa interactions are given by
 \bea
- {\cal L} &=&Y_{u2}\,\overline{Q}_L \, \tilde{{ \Phi}}_2 \,u_R
+\,Y_{d1}\,
\overline{Q}_L\,{\Phi}_1 \, d_R\, + \, Y_{\ell 1}\,\overline{L}_L \, {\Phi}_1\,e_R+\, \mbox{h.c.}\,, \eea where
$Q_L^T=(u_L\,,d_L)$, $L_L^T=(\nu_L\,,l_L)$,
$\widetilde\Phi_{1,2}=i\tau_2 \Phi_{1,2}^*$, and $Y_{u2}$,
$Y_{d1}$ and $Y_{\ell 1}$ are $3 \times 3$ matrices in family
space.

The Yukawa couplings of the neutral Higgs bosons normalized to the SM are given by
\bea\label{hffcoupling} &&
y^{h}_V=\sin(\beta-\alpha),~~~y_{f}^{h}=\left[\sin(\beta-\alpha)+\cos(\beta-\alpha)\kappa_f\right], \nonumber\\
&&y^{H}_V=\cos(\beta-\alpha),~~~y_{f}^{H}=\left[\cos(\beta-\alpha)-\sin(\beta-\alpha)\kappa_f\right], \nonumber\\
&&y^{A}_V=0,~~~y_{A}^{f}=-i\kappa_f~{\rm (for~u)},~~~~y_{f}^{A}=i \kappa_f~{\rm (for~d,~\ell)},\nonumber\\ 
&&{\rm with}~\kappa_d=\kappa_\ell\equiv-\tan\beta,~~~\kappa_u\equiv 1/\tan\beta,\eea 
where $\alpha$ is the mixing angle of the two CP-even Higgs bosons, and $V$ denotes $Z$ or $W$.

\subsection{A scalar dark matter}
 Now we add a real singlet scalar $S$ to the type-II 2HDM, and the potential 
containing the $S$ field is written as
\begin{eqnarray}
\mathcal{V}_{S}&=&{1\over 2}S^2(\kappa_{1}\Phi_1^\dagger \Phi_1
+\kappa_{2}\Phi_2^\dagger \Phi_2)+{m_{0}^2\over
2}S^2+{\lambda_S\over 4!}S^4\label{potent}.
\end{eqnarray} The linear and cubic terms of the $S$ field are
forbidden by a $Z'_2$ symmetry, under which $S\rightarrow -S$. The $S$ is a possible DM candidate provided
it does not acquire a VEV. We can obtain the DM mass and
the cubic interactions with the neutral Higgses from the Eq.
(\ref{potent}),
\begin{eqnarray}
m_S^2&=&m_0^2+\frac{1}{2}\kappa_1
v^2\cos^2\beta+\frac{1}{2}\kappa_2 v^2\sin^2\beta,\nonumber\\
-\lambda_{h} vS^2h/2&\equiv& -(-\kappa_{1}\sin\alpha\cos\beta+\kappa_{2}\cos\alpha\sin\beta)vS^2h/2,\nonumber\\
-\lambda_{H} vS^2H/2&\equiv&
-(\kappa_{1}\cos\alpha\cos\beta+\kappa_{2}\sin\alpha\sin\beta)vS^2H/2.
\label{dmcoup}\end{eqnarray}

\section{Numerical calculations}
In this paper, we discuss two different scenarios:

Case A: The light CP-even Higgs boson $h$ is taken as the 125 GeV Higgs, $m_h=125$ GeV, and $H$ and $A$ are heavier than 130 GeV.
Both $h$ and $H$ are the portals between the DM and SM sectors, and 
contribute to the DM interactions with SM particles.

Case B: The heavy CP-even Higgs boson $H$ is taken as the 125 GeV Higgs, $m_H=125$ GeV. The light CP-even Higgs $h$ is the
only portal between the DM and SM sectors, namely fixing $\lambda_H=0$. Thus, the invisible decay mode $H\to SS$ is absent, and
does not bring troubles to the signal data of the 125 GeV Higgs. The DM mass is slightly
below Higgs resonance, $m_h/2=(1.0\sim 1.2)\times m_S$. In the calculation of the thermal averaged cross section,
the kinetic energy of the DM is non negligible in the early universe, and as a result the resonant condition 
in the DM pair-annihilation can be met for $m_S$ slightly smaller than $m_h/2$. 
The temperature at the present time is much lower compared to the freeze-out temperature, and the velocity of DM is much smaller than
that in the early universe. Therefore, the resonant condition for the today DM pair-annihilation is hardly satisfied 
for $m_S$ slightly smaller than $m_h/2$.

In our calculations, to implement the constraints from the Higgs searches at the LHC,
we need employ $\textsf{SusHi}$ \cite{sushi} to compute cross sections of Higgs in the
gluon fusion and $b\bar{b}$-associated production at NNLO in QCD. Results of $\textsf{SusHi}$
might not be reasonable for a small Higgs mass. Therefore, we take $m_h>20$ GeV, which 
determines the DM mass to be larger than 10 GeV in the Case B.
The measurement of the branching fraction of $b \to s\gamma$ imposed the
strongest lower limit on the charged Higgs mass of type-II 2HDM, $m_{H^{\pm}} > 580$ GeV \cite{bsr580}.
The $S$, $T$ and $U$ oblique parameters give the stringent
constraints on the mass spectrum of Higgses of type II 2HDM \cite{1701.02678,1502.07532,1604.01406}. 
One of $m_A$ and $m_H$ is around 600 GeV, and another is allowed to have 
a wide mass range including low mass \cite{1701.02678}. Therefore, to allow $h$ to be light enough 
we fix $m_A=600$ GeV in the Case B.

In our calculations, we consider the following observables and constraints:
\begin{itemize}
\item[(1)] Theoretical constraints. The scalar potential of the model contains the 
potential type-II 2HDM and the potential of the DM sector. 
The vacuum stability, perturbativity, and tree-level unitarity impose constraints on the relevant parameters,
which are discussed in detail in Refs. \cite{2hisos-4,2hisos-6}. Here we employ the formulas in \cite{2hisos-4,2hisos-6} to
implement the theoretical constraints. Compared to Refs. \cite{2hisos-4,2hisos-6}, there are additional factors of $\frac{1}{2}$
in the $\kappa_1$ term and the $\kappa_2$ term of this paper.

\item[(2)] The oblique parameters. The $S$, $T$, $U$ parameters can impose strong constraints on 
the mass spectrum of Higgses of 2HDM. The $\textsf{2HDMC}$ \cite{2hc-1}
is employed to implement the constraints from
the oblique parameters ($S$, $T$, $U$).

\item[(3)] The flavor observables and $R_b$. $\textsf{SuperIso-3.4}$ \cite{spriso} is employed to 
consider the constraint of $B\to X_s\gamma$, and $\Delta m_{B_s}$ is calculated following the
formulas in \cite{deltmq}. Besides, we perform the constraints of bottom quarks produced in $Z$ decays, $R_b$,
which is calculated using the formulas in \cite{rb1,rb2}.

\item[(4)] The global fit to the signal data of the 125 GeV Higgs. 
Because the 125 GeV Higgs couplings with the SM particles in this model
can be modified compared to the SM, the SM-like decay modes will be corrected. 
In the Case A, $h$ is the 125 GeV Higgs, and the invisible decay $h\to SS$ is kinematically allowed,
which will be strongly constrained by the experimental data of the 125 GeV Higgs. 
In the Case B, $H$ is the 125 GeV Higgs, and the invisible decay $H\to SS$ is absent 
since the coupling $HSS$ is taken as zero. However, the decay $H\to hh$ is kinematically allowed for
$m_h<$ 62.5 GeV. We perform the
$\chi^2$ calculation for the signal strengths of the 125 GeV Higgs in the
$\mu_{ggF+tth}(Y)$ and $\mu_{VBF+Vh}(Y)$ with $Y$ denoting the decay
mode $\gamma\gamma$, $ZZ$, $WW$, $\tau^+ \tau^-$ and $b\bar{b}$,
 \begin{eqnarray} \label{eq:ellipse}
  \chi^2(Y) =\left( \begin{array}{c}
        \mu_{ggH+ttH}(Y) - \widehat{\mu}_{ggH+ttH}(Y)\\
        \mu_{VBF+VH}(Y) - \widehat{\mu}_{VBF+VH}(Y)
                 \end{array} \right)^T
                 \left(\begin{array}{c c}
                        a_Y & b_Y \\
                        b_Y & c_Y
                 \end{array}\right) \nonumber\\
\times
                  \left( \begin{array}{c}
        \mu_{ggH+ttH}(Y) - \widehat{\mu}_{ggH+ttH}(Y)\\
        \mu_{VBF+VH}(Y) - \widehat{\mu}_{VBF+VH}(Y)
                 \end{array} \right) \,.
 \end{eqnarray}
$\widehat{\mu}_{ggH+ttH}(Y)$ and $\widehat{\mu}_{VBF+VH}(Y)$
are the data best-fit values and $a_Y$, $b_Y$ and $c_Y$ are the
parameters of the ellipse, which are given by the
combined ATLAS and CMS experiments \cite{160602266}. We pay
particular attention to the surviving samples with
$\chi^2-\chi^2_{\rm min} \leq 6.18$, where $\chi^2_{\rm min}$
denotes the minimum of $\chi^2$. These samples correspond to be within
the $2\sigma$ range in any two-dimension plane of the
model parameters when explaining the Higgs data.

In addition, the ATLAS and CMS reported the upper limits on the branching ratio of 
invisible decay of the 125 GeV Higgs. In our calculation we impose the constraints,
Br$(h\to SS)<$ 0.34 \cite{160602266}.

\begin{table}
\begin{footnotesize}
\begin{tabular}{| c | c | c | c |}
\hline
\textbf{Channel} & \textbf{Experiment} & \textbf{Mass range (GeV)}  &  \textbf{Luminosity} \\
\hline
 {$gg/b\bar{b}\to \varphi'/A \to \tau^{+}\tau^{-}$} & ATLAS 8 TeV~\cite{47Aad:2014vgg} & 90-1000 & 19.5-20.3 fb$^{-1}$ \\
{$gg/b\bar{b}\to \varphi'/A \to \tau^{+}\tau^{-}$} & CMS 8 TeV~\cite{48CMS:2015mca} &  90-1000  &19.7 fb$^{-1}$ \\
{$gg/b\bar{b}\to \varphi'/A \to \tau^{+}\tau^{-}$} & ATLAS 13 TeV~\cite{82vickey} & 200-1200 &13.3 fb$^{-1}$ \\
{$gg/b\bar{b}\to \varphi'/A \to \tau^{+}\tau^{-}$} & CMS 13 TeV~\cite{add-hig-16-037} & 90-3200 &12.9 fb$^{-1}$ \\
{$gg\to \varphi'/A \to \tau^{+}\tau^{-}$} & CMS 13 TeV \cite{1709.07242}& 200-2250   & 36.1 fb$^{-1}$ \\
{$b\bar{b}\to \varphi'/A \to \tau^{+}\tau^{-}$} & CMS 13 TeV \cite{1709.07242}& 200-2250   & 36.1 fb$^{-1}$ \\
 {$b\bar{b}\to \varphi'/A \to \tau^{+}\tau^{-}$} & CMS 8 TeV \cite{1511.03610}& 20-80   & 19.7 fb$^{-1}$ \\
\hline
 {$b\bar{b}\to \varphi'/A \to \mu^{+}\mu^{-}$} & CMS 8 TeV~\cite{CMS-HIG-15-009} & 25-60 & 19.7 fb$^{-1}$ \\
\hline
 {$pp\to \varphi'/A \to \gamma\gamma$} & ATLAS 13 TeV \cite{80lenzi} & 200-2400 & 15.4 fb$^{-1}$ \\
{$gg\to \varphi'/A \to \gamma\gamma$}& CMS 8+13 TeV \cite{81rovelli}& 500-4000 & 12.9 fb$^{-1}$ \\
{$gg\to \varphi'/A \to \gamma\gamma$~+~$t\bar{t}\varphi'/A~(\varphi'/A\to \gamma\gamma)$}& CMS 8 TeV \cite{HIG-17-013-pas}& 80-110 & 19.7 fb$^{-1}$ \\
{$gg\to \varphi'/A \to \gamma\gamma$~+~$t\bar{t}\varphi'/A~(\varphi'/A\to \gamma\gamma)$}& CMS 13 TeV \cite{HIG-17-013-pas}& 70-110 & 35.9 fb$^{-1}$ \\
{$VV\to \varphi' \to \gamma\gamma$~+~$V\varphi'~(\varphi'\to \gamma\gamma)$}& CMS 8 TeV \cite{HIG-17-013-pas}& 80-110 & 19.7 fb$^{-1}$ \\
{$VV\to \varphi' \to \gamma\gamma$~+~$V\varphi'~(\varphi'\to \gamma\gamma)$}& CMS 13 TeV \cite{HIG-17-013-pas}& 70-110 & 35.9 fb$^{-1}$ \\
\hline

 {$gg/VV\to \varphi'\to W^{+}W^{-}$} & ATLAS 8 TeV  \cite{55Aad:2015agg}& 300-1500  &  20.3 fb$^{-1}$\\

{$gg/VV\to \varphi'\to W^{+}W^{-}~(\ell\nu\ell\nu)$} & ATLAS 13 TeV  \cite{77atlasww13}& 300-3000  &  13.2 fb$^{-1}$\\

{$gg\to \varphi'\to W^{+}W^{-}~(\ell\nu qq)$} & ATLAS 13 TeV  \cite{78atlasww13lvqq}& 500-3000  &  13.2 fb$^{-1}$\\

{$gg/VV\to \varphi'\to W^{+}W^{-}~(\ell\nu qq)$} & ATLAS 13 TeV  \cite{1710.07235}& 200-3000  &  36.1 fb$^{-1}$\\
{$gg/VV\to \varphi'\to W^{+}W^{-}~(e\nu \mu\nu)$} & ATLAS 13 TeV  \cite{1710.01123}& 200-3000  &  36.1 fb$^{-1}$\\
\hline

$gg/VV\to \varphi'\to ZZ$ & ATLAS 8 TeV \cite{57Aad:2015kna}& 160-1000 & 20.3 fb$^{-1}$ \\

$gg\to \varphi' \to ZZ(\ell \ell \nu \nu)$ & ATLAS 13 TeV~\cite{74koeneke4} & 300-1000  & 13.3 fb$^{-1}$ \\
$gg\to \varphi'\to ZZ(\nu \nu qq)$ & ATLAS 13 TeV~\cite{75koeneke5} & 300-3000 & 13.2 fb$^{-1}$ \\
$gg/VV\to \varphi'\to ZZ(\ell \ell qq)$ & ATLAS 13 TeV~\cite{75koeneke5} & 300-3000 & 13.2 fb$^{-1}$ \\
$gg/VV\to \varphi'\to ZZ(\ell\ell\ell\ell)$ & ATLAS 13 TeV~\cite{76koeneke3} & 200-3000 & 14.8 fb$^{-1}$ \\

$gg/VV\to \varphi'\to ZZ(\ell\ell\ell\ell+\ell\ell\nu\nu)$ & ATLAS 13 TeV~\cite{1712.06386} & 200-2000 & 36.1 fb$^{-1}$ \\
$gg/VV\to \varphi'\to ZZ(\nu\nu qq+\ell\ell qq)$ & ATLAS 13 TeV~\cite{1708.09638} & 300-5000 & 36.1 fb$^{-1}$ \\
\hline

\end{tabular}
\end{footnotesize}
\caption{The upper limits at 95\%  C.L. on the production cross-section times branching ratio of
$\tau^+\tau^-$, $\mu^+\mu^-$, $\gamma\gamma$, $WW$ and $ZZ$ considered in 
the $\varphi'$ and $ A $ searches at the LHC. Here $\varphi'$ denotes the non-SM-like CP-even Higgs in 
2HDM.}
\label{tabh}
\end{table}

\begin{table}
\begin{footnotesize}
\begin{tabular}{| c | c | c | c |}
\hline
\textbf{Channel} & \textbf{Experiment} & \textbf{Mass range (GeV)}  &  \textbf{Luminosity} \\
\hline
$gg\to\varphi'\to \varphi_{s}\varphi_{s} \to (\gamma \gamma) (b \bar{b})$ & CMS 8 TeV \cite{64Khachatryan:2016sey} & 250-1100  & 19.7 fb$^{-1}$\\

$gg\to\varphi'\to \varphi_{s}\varphi_{s} \to (b\bar{b}) (b\bar{b})$ & CMS 8 TeV \cite{65Khachatryan:2015yea}&   270-1100   & 17.9 fb$^{-1}$\\

$gg\to\varphi'\to \varphi_{s}\varphi_{s} \to (b\bar{b}) (\tau^{+}\tau^{-})$ & CMS 8 TeV \cite{66Khachatryan:2015tha}&  260-350 & 19.7 fb$^{-1}$\\

$gg\to\varphi'\to \varphi_{s}\varphi_{s} \to (\gamma \gamma) (b \bar{b})$ & ATLAS 13 TeV \cite{add-CONF-2016-004} & 275-400  & 3.2 fb$^{-1}$\\

$gg\to\varphi'\to \varphi_{s}\varphi_{s} \to (\gamma \gamma) (b \bar{b})$ & CMS 13 TeV \cite{add-hig-16-032} & 250-900  & 2.7 fb$^{-1}$\\

$gg \to\varphi'\to \varphi_{s}\varphi_{s} \to b\bar{b}b\bar{b}$ & ATLAS 13 TeV~\cite{84varol} & 300-3000  &  13.3 fb$^{-1}$ \\

$gg \to\varphi'\to \varphi_{s}\varphi_{s} \to (b\bar{b}) (\tau^{+} \tau^{-})$ & CMS 13 TeV~\cite{85CMS:2016knm} & 250-900  &  12.9 fb$^{-1}$ \\

$gg \to\varphi'\to \varphi_{s}\varphi_{s} \to b\bar{b}b\bar{b}$ & CMS 13 TeV~\cite{1710.04960} & 750-3000  &  35.9 fb$^{-1}$ \\
$gg \to\varphi'\to \varphi_{s}\varphi_{s} \to (b\bar{b}) (\tau^{+}\tau^{-})$ & CMS 13 TeV~\cite{1707.02909} & 250-900  &  35.9 fb$^{-1}$ \\
$gg \to\varphi'\to \varphi_{s}\varphi_{s} \to (WW^*) (\gamma\gamma)$ & ATLAS 13 TeV~\cite{1707.02909} & 260-500  &  13.3 fb$^{-1}$ \\
\hline

$gg\to A\to \varphi_{s}Z \to (\tau^{+}\tau^{-}) (\ell \ell)$ & CMS 8 TeV \cite{66Khachatryan:2015tha}& 220-350 & 19.7 fb$^{-1}$\\

$gg\to A\to \varphi_{s}Z \to (b\bar{b}) (\ell \ell)$ & CMS 8 TeV \cite{67Khachatryan:2015lba} & 225-600 &19.7 fb$^{-1}$ \\

$gg\to A\to \varphi_{s}Z\to (\tau^{+}\tau^{-}) Z$ & ATLAS 8 TeV \cite{68Aad:2015wra}&220-1000 & 20.3 fb$^{-1}$ \\

 {$gg\to A\to \varphi_{s}Z\to (b\bar{b})Z$} & ATLAS 8 TeV \cite{68Aad:2015wra}& 220-1000 & 20.3 fb$^{-1}$  \\
{$gg/b\bar{b}\to A\to \varphi_{s}Z\to (b\bar{b})Z$}& ATLAS 13 TeV \cite{69AZhatlas13}& 200-2000 & 3.2 fb$^{-1}$  \\

{$gg/b\bar{b}\to A\to \varphi_{s}Z\to (b\bar{b})Z$}& ATLAS 13 TeV \cite{1712.06518}& 200-2000 & 36.1 fb$^{-1}$  \\
\hline

 {$gg\to \varphi_{s} \to AA/\varphi'\varphi' \to \tau^{+}\tau^{-}\tau^{+}\tau^{-}$} & ATLAS 8 TeV~\cite{1505.01609} & 4-50 & 20.3 fb$^{-1}$ \\
{$pp\to  \varphi_{s} \to AA/\varphi'\varphi' \to \tau^{+}\tau^{-}\tau^{+}\tau^{-}$} & CMS 8 TeV~\cite{1701.02032} &  5-15  &19.7 fb$^{-1}$ \\
{$pp\to  \varphi_{s} \to AA/\varphi'\varphi' \to (\mu^{+}\mu^{-})(b\bar{b})$} & CMS 8 TeV~\cite{1701.02032} &  25-62.5  &19.7 fb$^{-1}$ \\
{$pp\to  \varphi_{s} \to AA/\varphi'\varphi' \to (\mu^{+}\mu^{-})(\tau^{+}\tau^{-})$} & CMS 8 TeV~\cite{1701.02032} &  15-62.5  &19.7 fb$^{-1}$ \\
\hline
$gg\to A(\varphi')\to \varphi'(A)Z\to (b\bar{b}) (\ell \ell)$ & CMS 8 TeV \cite{160302991} & 40-1000 &19.8 fb$^{-1}$ \\

$gg\to A(\varphi')\to \varphi'(A)Z\to (\tau^{+}\tau^{-}) (\ell \ell)$ & CMS 8 TeV \cite{160302991}& 20-1000 & 19.8 fb$^{-1}$ \\
\hline
\end{tabular}
\end{footnotesize}
\caption{The upper limits at 95\%  C.L. on the production cross-section times branching ratio for the channels
of Higgs-pair and a Higgs production in association with $Z$ at the LHC. Here $\varphi'$ and $\varphi_{s}$
denote the non-SM-like CP-even Higgs and the 125 GeV Higgs in the 2HDM, respectively.}
\label{tabhh}
\end{table}

\item[(5)] The non-observation of additional Higgs bosons. We employ
$\textsf{HiggsBounds}$ \cite{hb1,hb2} to implement the exclusion
constraints from the searches for the neutral and charged Higgs at the LEP at 95\% confidence level.
Especially for the Case B, the searches for a light Higgs at the LEP can impose stringent constraints
on the parameter space.

At the LHC, the ATLAS and CMS have searched for an additional scalar via its decay into various SM channels and
some exotic decays. For $gg\to A$ production in type-II 2HDM, the contributions
of $b$-quark loop interfere destructively with those of top quark loop. 
The cross section decreases with an increasing
of $\tan\beta$, reaches the minimum value for the moderate value of
$\tan\beta$, and is dominated by the $b$-quark loop for enough large
value of $\tan\beta$. The cross section of the CP-even Higgs in the gluon fusion depends on $\sin(\beta-\alpha)$
in addition to the Higgs mass and $\tan\beta$.
 We use $\textsf{SusHi}$ \cite{sushi} to compute cross sections for Higgs in the
gluon fusion and $b\bar{b}$-associated production at NNLO in QCD. A complete list of the searches for
additional Higgs considered by us is summarized in 
Table \ref{tabh} and Table \ref{tabhh} where some channels are taken from Ref. \cite{1608.02573}. 
Refs. \cite{mhp500,1706.07414} show that the LHC searches for the charged Higgs fail to constrain the
model for $m_{H^{\pm}}>500$ GeV. Therefore, the searches channels of the charged Higgs are not
included in this paper.

\item[(6)] The DM observables. We use $\textsf{micrOMEGAs}$ \cite{micomega} 
to calculate the relic density and today DM pair-annihilation.
The model file is generated by $\textsf{FeynRules}$ \cite{feyrule}.
For 10 GeV $<m_S<$ 50 GeV, the DM will annihilate into $b\bar{b}$ dominantly in this model.

In this model, the elastic scattering of $S$ on a nucleon receives
the contributions of the process with $t-$ channel exchange of $h$ and $H$ in the Case A, and only 
$h$ exchange in the Case B. If both $h$ and $H$ contribute to the DM interactions with nucleons, 
the spin-independent cross section is given by
\cite{sigis},
 \beq \sigma_{p(n)}=\frac{\mu_{p(n)}^{2}}{4\pi m_{S}^{2}}
    \left[f^{p(n)}\right]^{2},
\eeq where $\mu_{p(n)}=\frac{m_Sm_{p(n)}}{m_S+m_{p(n)}}$, \beq
f^{p(n)}=\sum_{q=u,d,s}f_{q}^{p(n)}\mathcal{C}_{S
q}\frac{m_{p(n)}}{m_{q}}+\frac{2}{27}f_{g}^{p(n)}\sum_{q=c,b,t}\mathcal{C}_{S
q}\frac{m_{p(n)}}{m_{q}},\label{fpn} \eeq with $\mathcal{C}_{S
q}=\frac{\lambda_h}{m_h^2} m_q y_q^h + \frac{\lambda_H}{m_H^2} m_q y_q^H$. 
The values of the form factors $f_{q}^{p,n}$ and $f_{g}^{p,n}$ are extracted from $\textsf{micrOMEGAs}$ \cite{micomega}

Recently, the Planck collaboration reported the density of cold DM in the universe,
 $\Omega_{c}h^2 = 0.1198 \pm 0.0015$ \cite{planck}. 
The PandaX-II (2017) and the XENON1T (2017) respectively impose the strongest constraints on 
the spin-independent DM-nucleon cross section for $m_S>100$ GeV and $m_S<60$ GeV \cite{pandax,xenon}. 
The upper limits of PandaX-II (2017) are nearly the same as those of XENON1T (2017) for the DM with a mass 
of $60 \sim 100$ GeV. The Fermi-LAT searches for the DM annihilation from dSphs gave the upper limits on the averaged cross sections
of the DM annihilation to $e^+ e^-$, $\mu^+ \mu^-$, $\tau^+\tau^-$, $u\bar{u}$, $b\bar{b}$, and $WW$ \cite{fermi}.

\end{itemize}

\section{Results and discussions}
\subsection{Case A}

\begin{figure}[tb]
 \epsfig{file=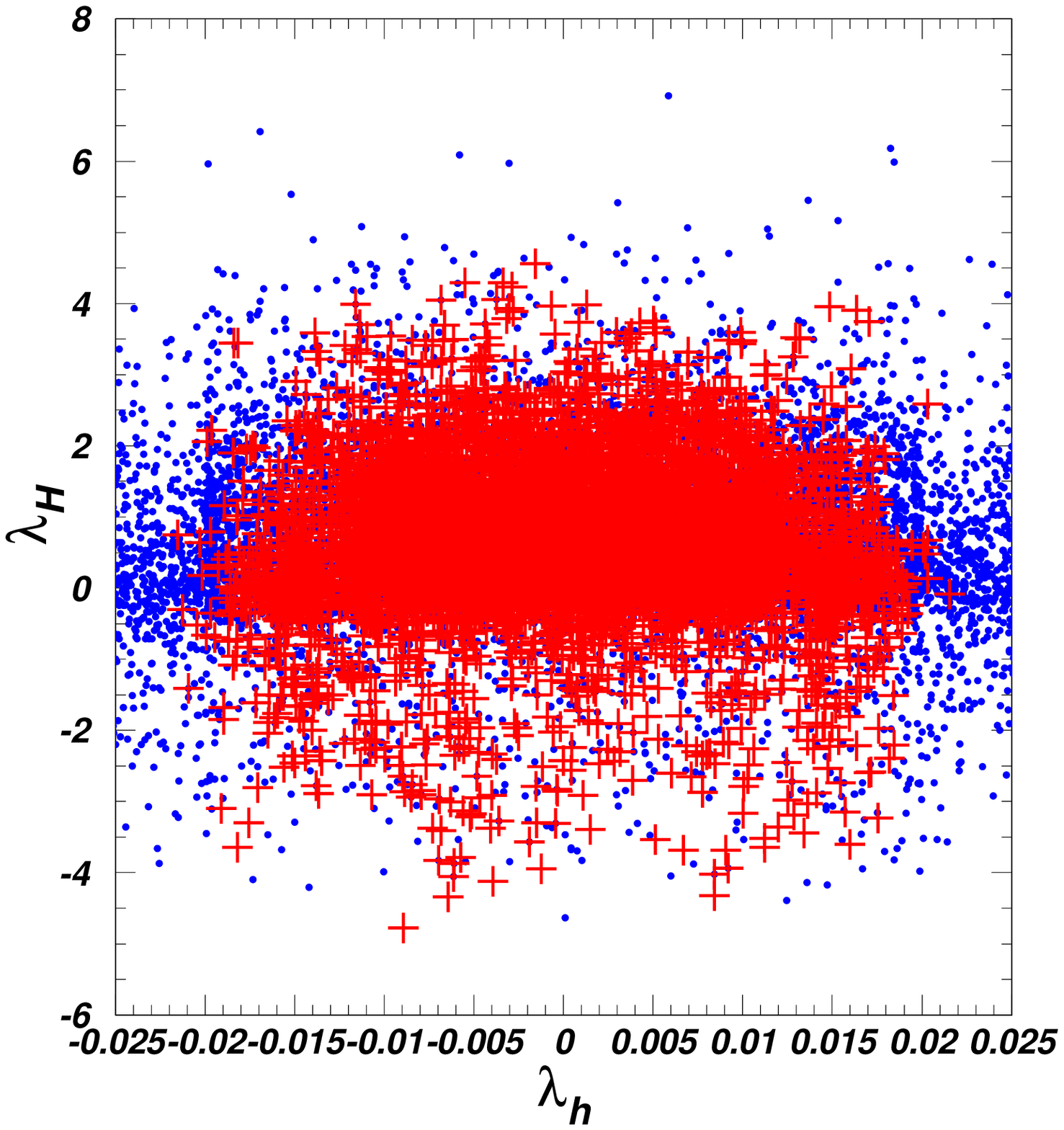,height=7.3cm}
  \epsfig{file=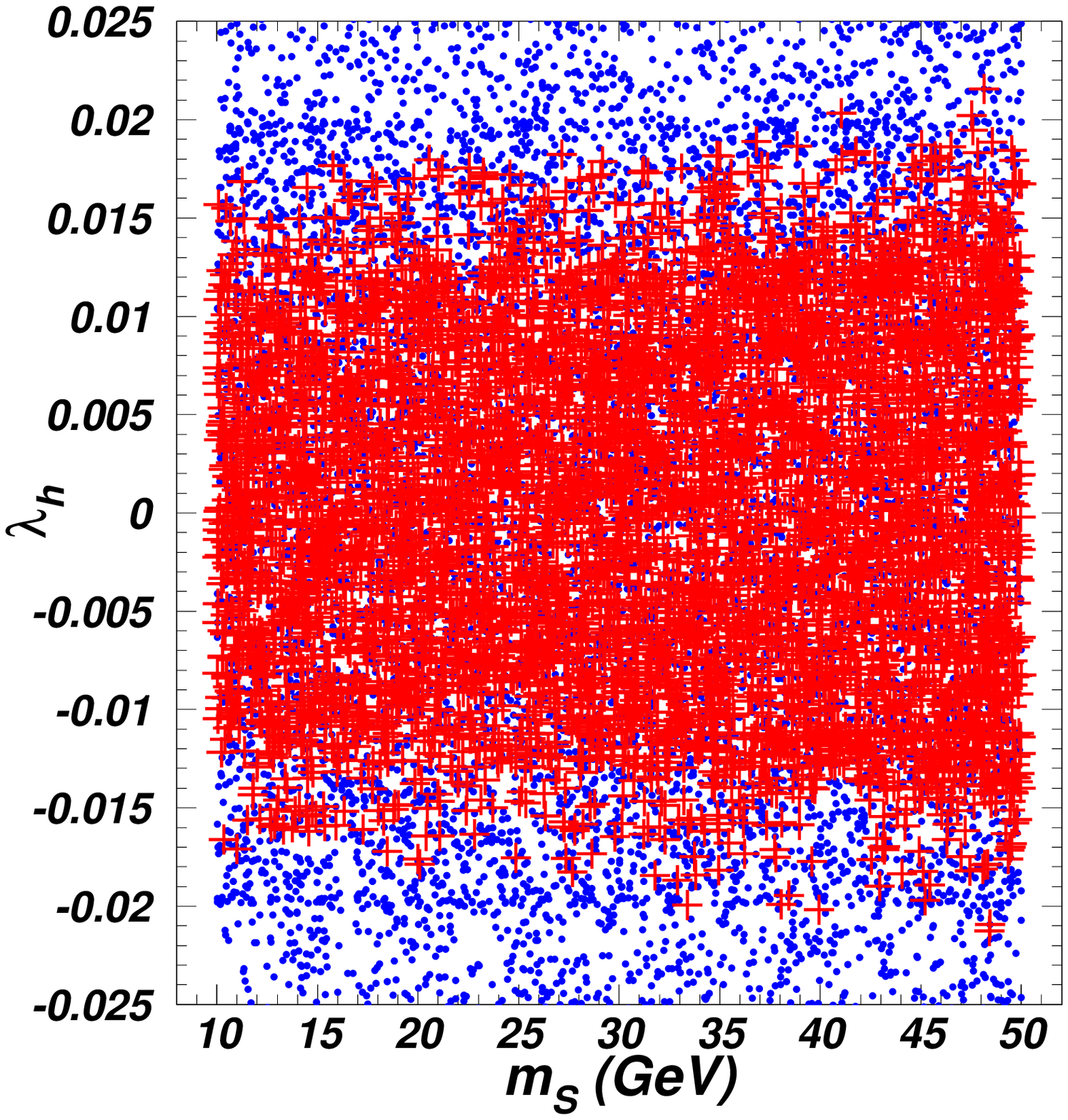,height=7.3cm}
\vspace{-0.5cm} \caption{In the Case A, the surviving samples projected on the
planes of $\lambda_H$ versus $\lambda_h$ and $\lambda_h$ versus
$m_{S}$. All the samples are allowed by the constraints from the vacuum stability, perturbativity, unitarity
 and the oblique parameters. The pluses (red) are also allowed 
by the joint constraints from the signal data of the 125 GeV Higgs, the flavor observables, $R_b$, and
 the exclusion limits from searches for Higgs at LEP.} \label{figew}
\end{figure}

\begin{figure}[tb]
  \epsfig{file=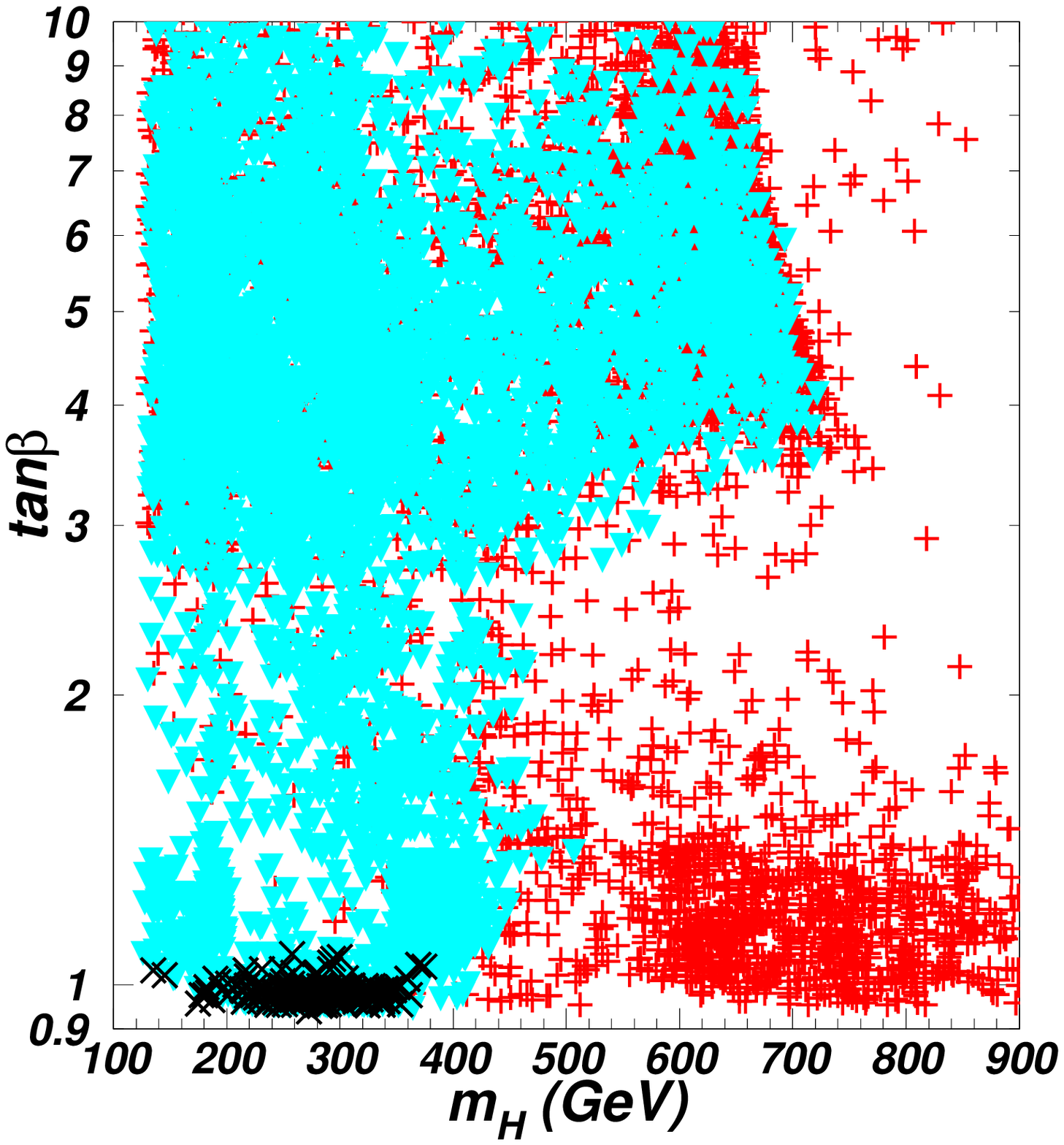,height=7.3cm}
  \epsfig{file=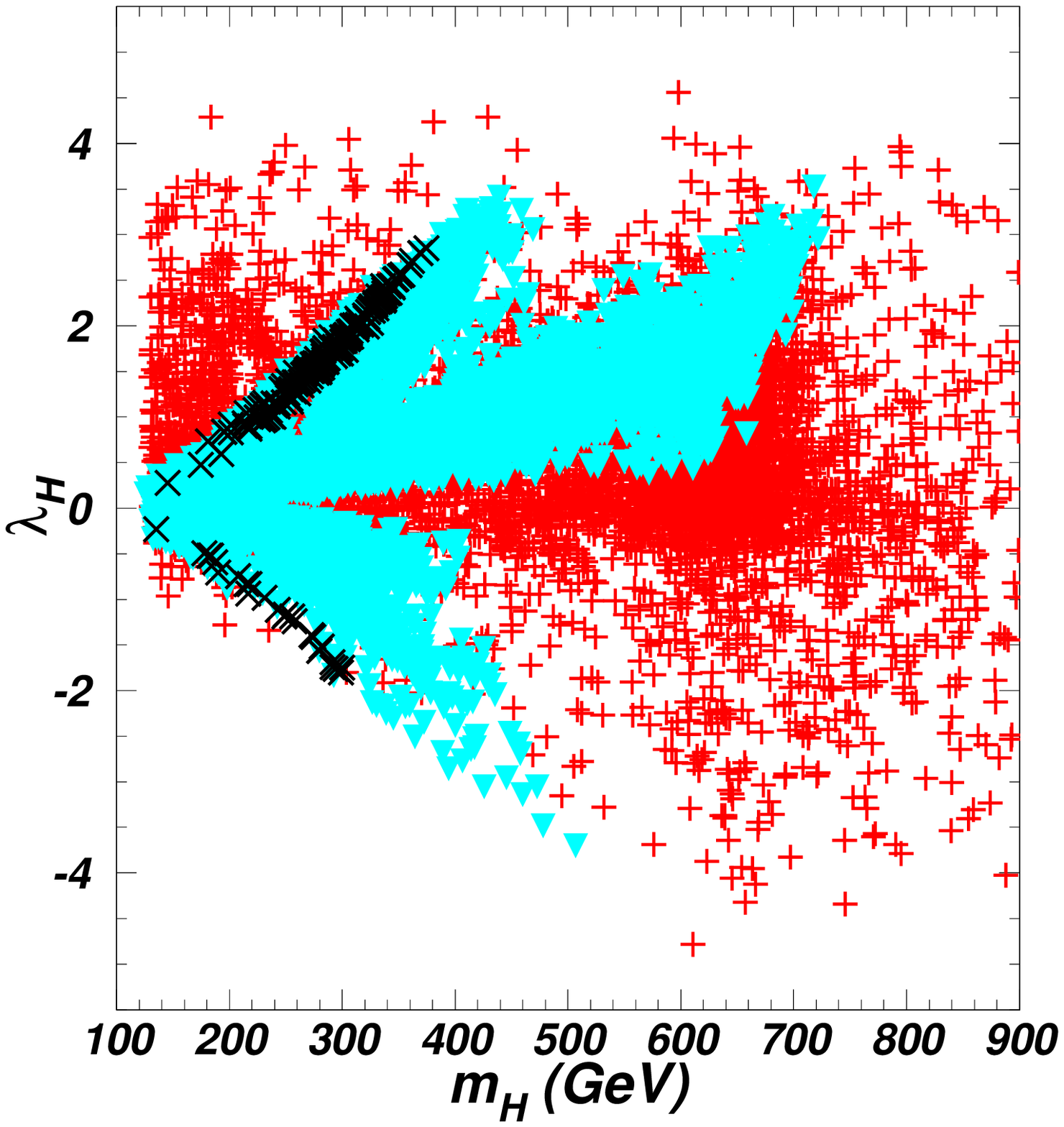,height=7.3cm}
\vspace{-0.5cm} \caption{In the Case A, the surviving samples projected on the
planes of $\tan\beta$ versus $m_H$ and $\lambda_H$ versus $m_H$.
All the samples are allowed by the constraints of "pre-LHC" and the signal data of the 125 GeV Higgs.
Also the inverted triangles (sky blue) are allowed by the relic density, and the crosses (black) are allowed by 
the relic density, the XENON1T (2017), and PandaX-II (2017).}
\label{figxenon}
\end{figure}

\begin{figure}[tb]
   \epsfig{file=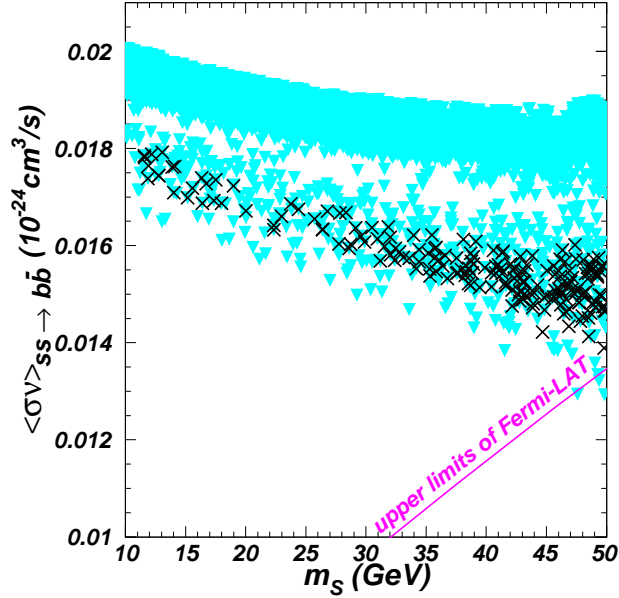,height=8cm}
 \vspace{-.3cm}
   \caption{In the Case A, the surviving samples projected on the
planes of $<\sigma v>_{SS\to b\bar{b}}$ versus $m_S$. The meanings of the inverted triangles (sky blue)
and crosses (black) are the same as Fig. \ref{figxenon}.} \label{fermi}
 \end{figure}

\begin{figure}[tb]
  \epsfig{file=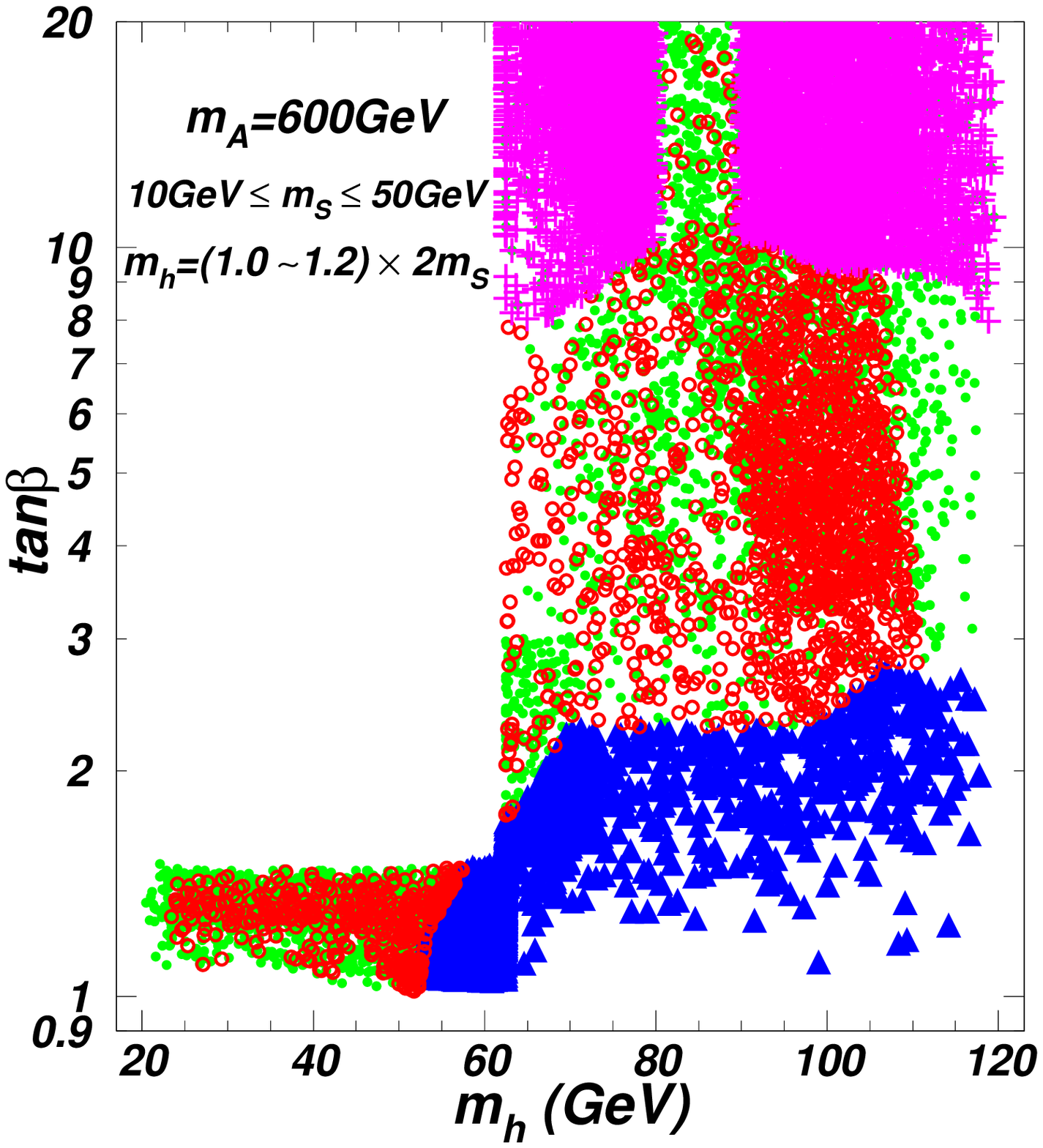,height=7.3cm}
  \epsfig{file=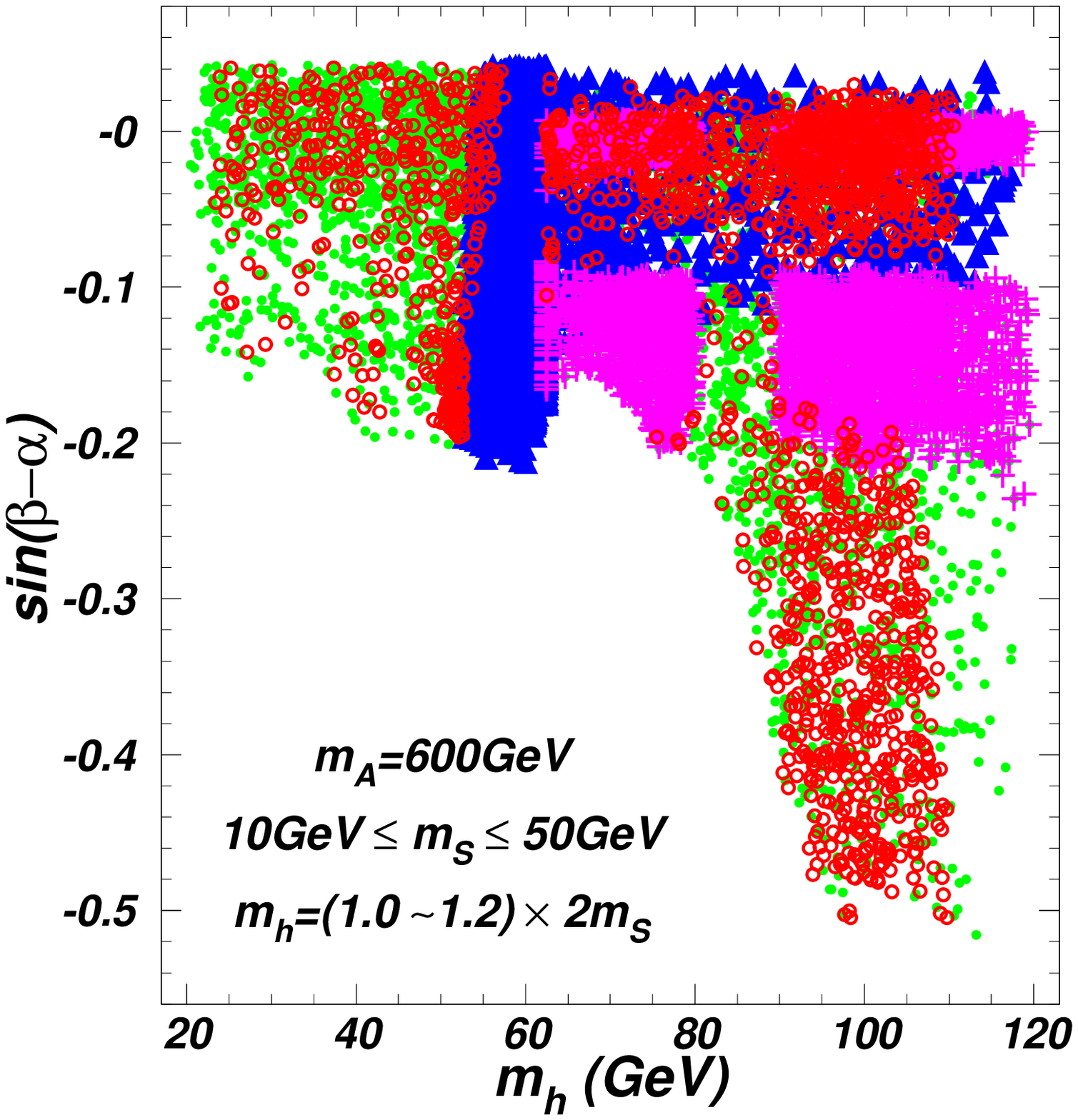,height=7.3cm}
\vspace{-0.5cm} \caption{In the Case B, the surviving samples projected on the
planes of $\tan\beta$ versus $m_h$ and $\sin(\beta-\alpha)$ versus $m_h$.
All the samples are allowed by the constraints of "pre-LHC" and the signal data of the 125 GeV Higgs.
The pluses (pink) and the triangles (royal blue) are respectively excluded by the $h\to \tau^+\tau^-$ and 
$A\to hZ$ channels at the LHC, and the other points are allowed by the two channels. 
The circles (red) are also allowed by the constraints of  
the relic density, XENON1T (2017), PandaX-II (2017) and the Fermi-LAT.}
\label{bfigmh}
\end{figure}

\begin{figure}[tb]
    \epsfig{file=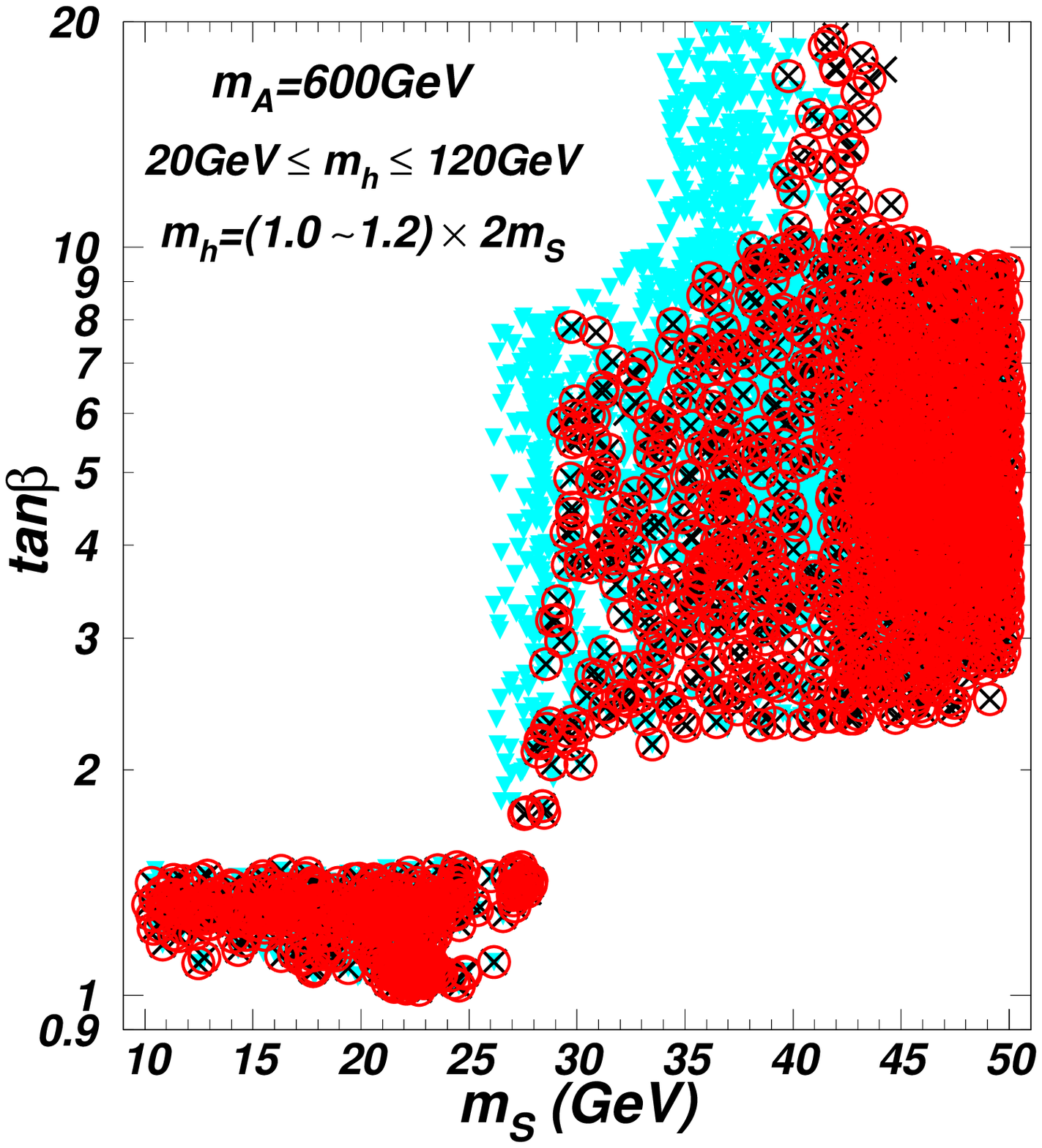,height=5.625cm}
  \epsfig{file=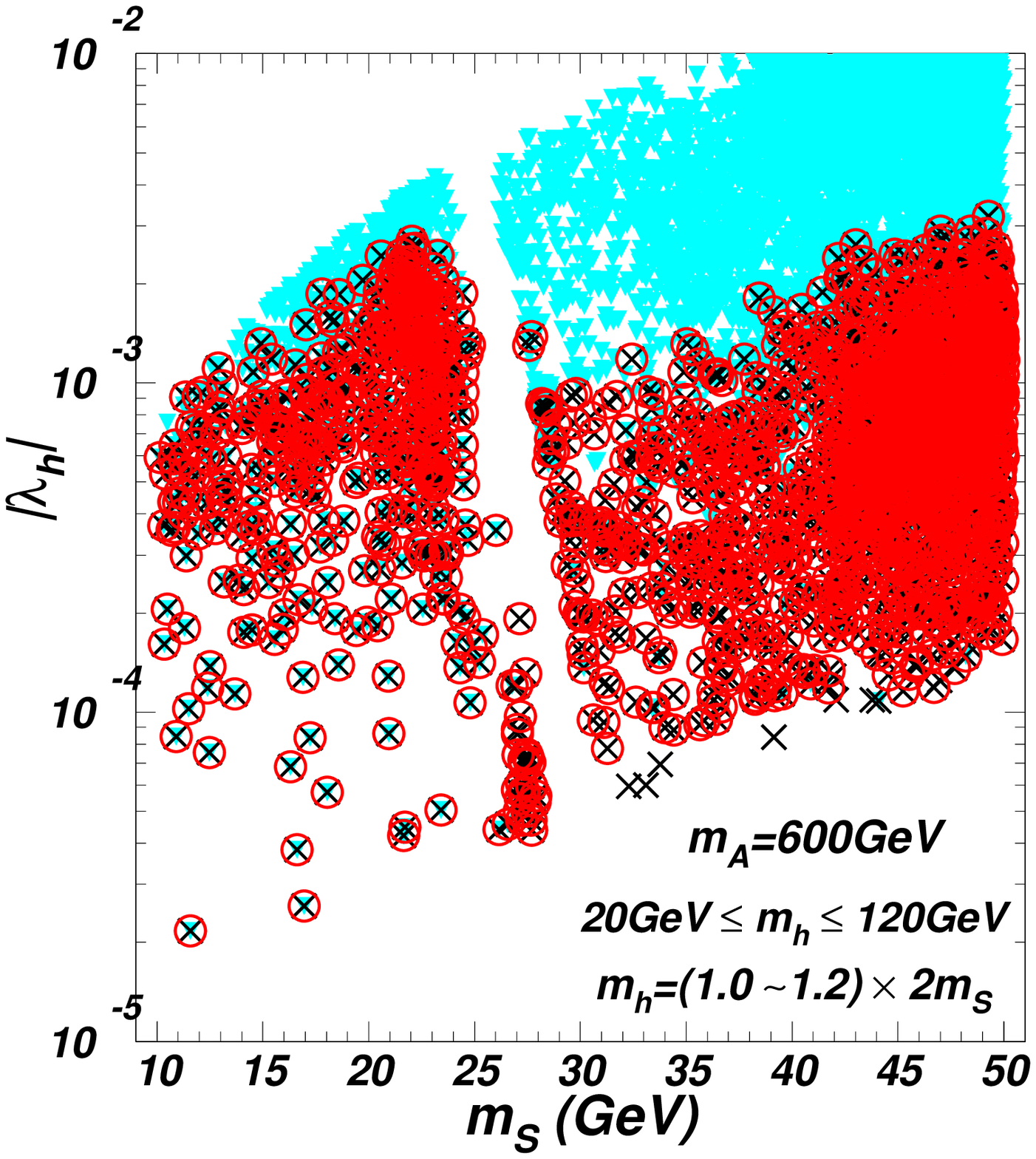,height=5.81cm}
  \epsfig{file=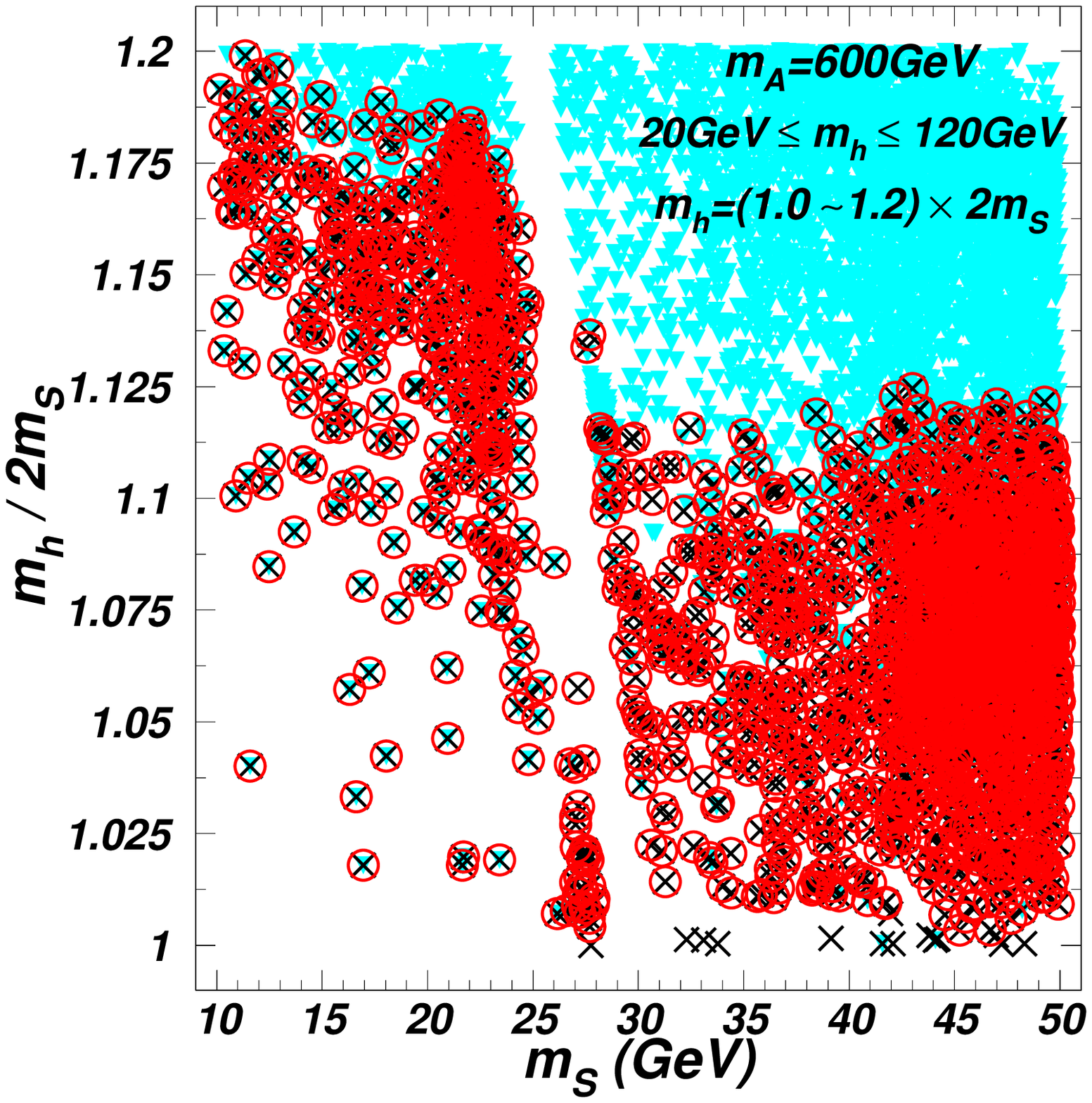,height=5.59cm}
\vspace{-0.25cm} \caption{In the Case B, the surviving samples projected on the
planes of $\tan\beta$ versus $m_S$, $\lambda_h$ versus $m_S$, and $\frac{m_h}{2m_S}$ versus $m_S$. 
All the samples are allowed by the constraints of "pre-LHC", the signal data of the 125 GeV Higgs, and the relic density.
Also the crosses (black) are allowed by the XENON1T (2017) and PandaX-II (2017),
and the circles (red) are allowed by the XENON1T (2017), PandaX-II (2017), and the Fermi-LAT.}
\label{bfigms}
\end{figure}

In Fig. \ref{figew}, we show $\lambda_H$ and $\lambda_h$ in the Case A allowed by the constraints of theory, the oblique parameters,
the signal data of the 125 GeV Higgs, the flavor observables, $R_b$, and
 the exclusion limits from searches for Higgs at LEP. The left panel shows that the vacuum stability, perturbativity, unitarity
 and the oblique parameters impose upper and lower limits on $\lambda_H$, -5 $<\lambda_H<$ 7.
Because the invisible decay $h\to SS$ is kinematically allowed, the signal data of the 125 GeV Higgs impose 
strong upper limits on $\mid\lambda_h\mid$, $\mid\lambda_h\mid <$ 0.017 (0.022) for $m_S$= 10 (50) GeV.

In Fig. \ref{figxenon}, we project the surviving samples on the
planes of $\tan\beta$ versus $m_H$ and $\lambda_H$ versus $m_H$ after imposing the constraints of "pre-LHC"
(denoting the theory, the oblique parameters, the flavor observables, $R_b$, and
 the exclusion limits from searches for Higgs at LEP), the signal data of the 125 GeV Higgs,
the relic density, XENON1T (2017), and PandaX-II (2017). Because the signal data of the 125 GeV Higgs
impose the strong upper limits on $\mid\lambda_h\mid$, the DM interactions mediated by $H$ play a key role in
the relic density. From Fig. \ref{figxenon}, we find that the model can give the correct relic density 
for the heavy CP-even Higgs mass up to 700 GeV. With an increase of $m_H$, $\lambda_H$ and $\tan\beta$ is favored to have large values,
which can enhance the couplings of $HSS$ and $Hb\bar{b}$. This is because
the $H$-mediated annihilation amplitude is suppressed by its large mass, and the
$H$ couplings to the DM and the SM particles are required to be large enough to obtain the correct annihilation rate.
Most of the parameter space are excluded by the PandaX-II (2017) and the XENON1T (2017), and only the narrow regions
 of $M_H<$ 400 GeV and $\tan\beta$ around 1.0 are allowed. For $\tan\beta$ around 1 and $\mid\sin(\beta-\alpha)\mid$ close to 1,
$y_d^H/y_u^H$ approaches to -1. For such a case, the DM interactions with nucleons mediated by $H$ can have a large isospin violation,
which can weaken the bounds of the PandaX-II and the XENON1T sizably. Because $\tan\beta$ is restricted to be around 1.0, 
an appropriate value of $\lambda_H$ is required to obtain the correct relic density, 
such as $\mid\lambda_H\mid$ around 1.8 for $m_H=$ 300 GeV, as shown in the right panel.

The Fig. \ref{fermi} shows that the limits of the Fermi-LAT searches for DM annihilation from dSphs exclude the whole
region of 10 GeV $<m_S<50$ GeV, including the parameter space surviving from the PandaX-II (2017) and the XENON1T (2017) bounds.
With the decreasing of $m_S$, the values of the today DM pair-annihilation into $b\bar{b}$ in this model 
exceed the Fermi-LAT upper limits sizably.

\subsection{Case B}
Now we discuss the Case B in which the heavy CP-even Higgs is the 125 GeV Higgs.
In Fig. \ref{bfigmh}, we project the surviving samples on the
planes of $\tan\beta$ versus $m_h$ and $\sin(\beta-\alpha)$ versus $m_h$ after imposing the constraints of "pre-LHC", 
the signal data of the 125 GeV Higgs, the searches for the additional Higgs at LHC, the DM relic density,
XENON1T (2017), PandaX-II (2017), and the Fermi-LAT searches for DM
annihilation from dSphs. The left panel shows that the signal data of the 125 GeV Higgs
restrict $\tan\beta$ to be in the range of $1\sim 1.5$ for $m_h<$ 62 GeV. For such range of $m_h$, 
the decay $H\to hh$ is kinematically open, and
enhance the total width of the 125 GeV Higgs. Since the width of $H\to hh$ is strongly constrained,
the searches for a light Higgs via $H\to hh$ channels at the LHC fail to impose  constraints on the
parameter space. The right panel shows that $\sin(\beta-\alpha)$ is imposed a lower bound for a given value of $m_h$.
This is because the $hZZ$ coupling is proportional to $\sin(\beta-\alpha)$, and 
a large absolute value of $\sin(\beta-\alpha)$ is excluded by the searches for a light Higgs via $e^+e^-\to Zh$ at the LEP.

The left panel of Fig. \ref{bfigmh} shows that the $gg\to A \to hZ$ channels at the LHC impose lower bounds on $\tan\beta$ for
53 GeV $<m_h<120$ GeV with $m_A$ being taken as 600 GeV, such as $\tan\beta>1.3$ for $m_h=$ 55 GeV, $\tan\beta>2.3$ for $m_h=$ 70 GeV, and 
$\tan\beta>2.7$ for $m_h=$ 110 GeV. The $AhZ$ coupling is proportional to $\cos(\beta-\alpha)$ and as a result
the decay $A\to hZ$ is not suppressed by $\sin(\beta-\alpha)$. However, the cross section
of $gg\to A$ will sizably decrease with an increasing of $\tan\beta$. 
The $b\bar{b}\to h \to \tau^+\tau^-$ channels impose upper limits on $\tan\beta$, and
$\tan\beta> 10$ is excluded for both $m_h<80$ GeV and $m_h>90$ GeV. In the range of 80 GeV $<m_h<90$ GeV, there is no available
experimental data of $b\bar{b}\to h \to \tau^+\tau^-$ from the ATLAS and CMS.
 
 Fig. \ref{bfigmh} shows that a narrow region of $m_h$ around 60 GeV and $m_A=600$ GeV 
is excluded by the joint constraints of the 125 GeV Higgs signal data and the $gg\to A \to hZ$ channels at the LHC.
In the other region of $m_h$, the DM with a mass of $10\sim 50$ GeV is allowed by 
the constraints of the relic density, XENON1T (2017), PandaX-II (2017), and the Fermi-LAT searches for DM
annihilation from dSphs. Certainly, the DM coupling with $h$ will play an important role.

In Fig. \ref{bfigms}, we project the surviving samples on the
planes of $\tan\beta$ versus $m_S$, $\lambda_h$ versus $m_S$, and $\frac{m_h}{2m_S}$ versus $m_S$.
From Fig. \ref{bfigms}, we find that for appropriate values of $\tan\beta$, $\lambda_h$ and $m_h$, 
the DM with a mass of $10\sim 50$ GeV is allowed by the constraints of "pre-LHC", 
the signal data of the 125 GeV Higgs, the searches for the additional Higgs at LHC, the DM relic density,
XENON1T (2017), PandaX-II (2017), and the Fermi-LAT. The right panel shows that 
the XENON1T (2017) and PandaX-II (2017) exclude the region of $\frac{m_h}{2m_S}>$ 1.125 and 30 GeV $<m_S<$ 50 GeV. In such range,
the kinetic energy of DM in the early universe can not offset the splitting of $m_h$ and $2m_S$, and the
resonant condition in the DM pair-annihilation is not met. Therefore, a large $hSS$ coupling is required to
obtain the correct relic density, and leads the spin-independent DM-nucleon cross section to 
exceed the upper limits of the XENON1T (2017) and PandaX-II (2017). Several points with $\frac{m_h}{2m_S}$ very close to 1.0
are allowed by the XENON1T (2017) and PandaX-II (2017), but excluded by the Fermi-LAT limits. 
This is because the resonant condition for the today DM pair-annihilation is also satisfied for $\frac{m_h}{2m_S}$ very close to 1.0.
The left panel shows that $\tan\beta$ is restricted to be in
the range of $1.0\sim1.5$ for 10 GeV $<m_s<$ 26 GeV. The middle panel shows that $\mid\lambda_h\mid$
is allowed to be as low as $10^{-5}$ due to the $h$ resonance contributions to the DM pair-annihilation.

\begin{figure}[tb]
  \epsfig{file=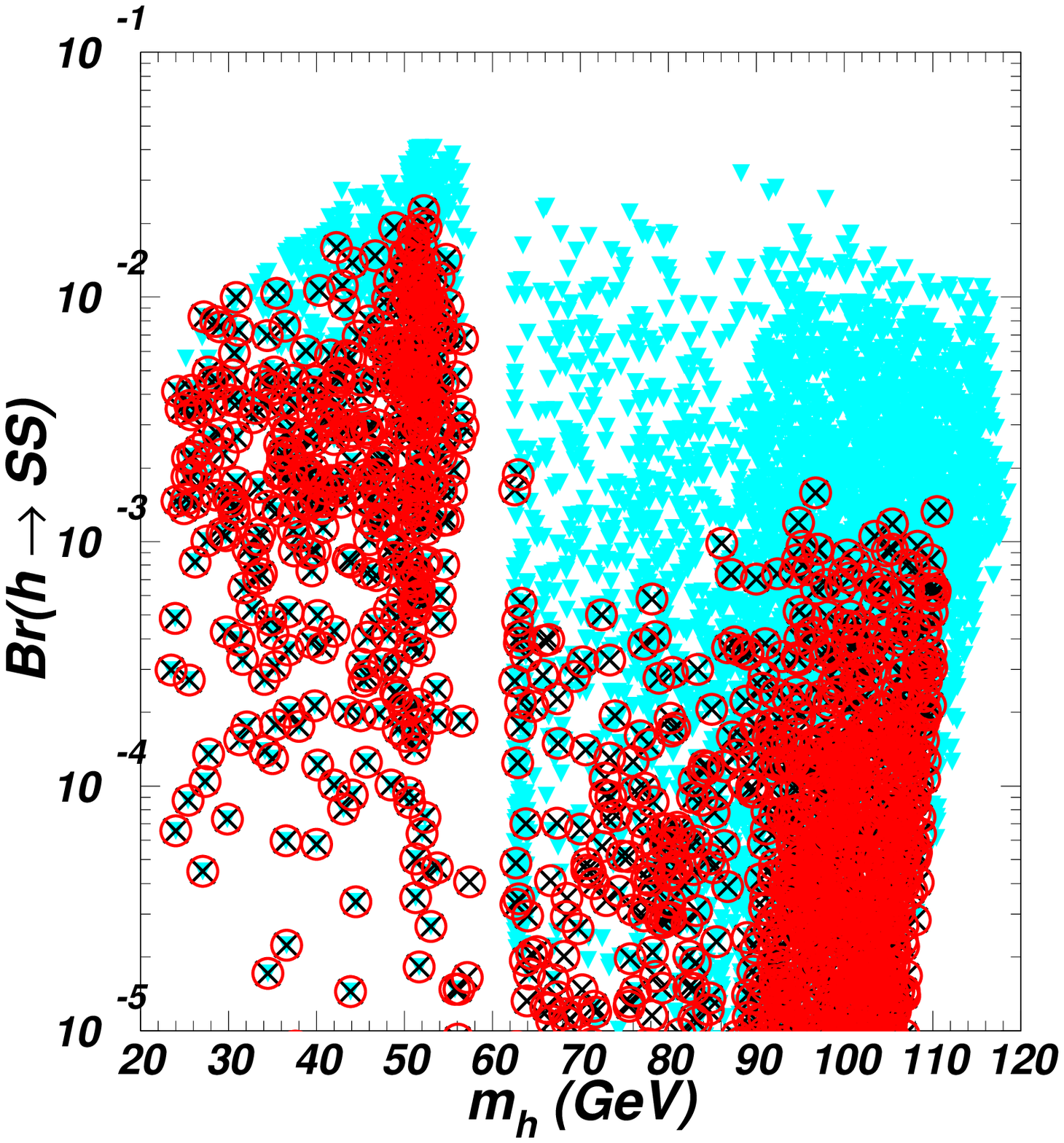,height=7.3cm}
  \epsfig{file=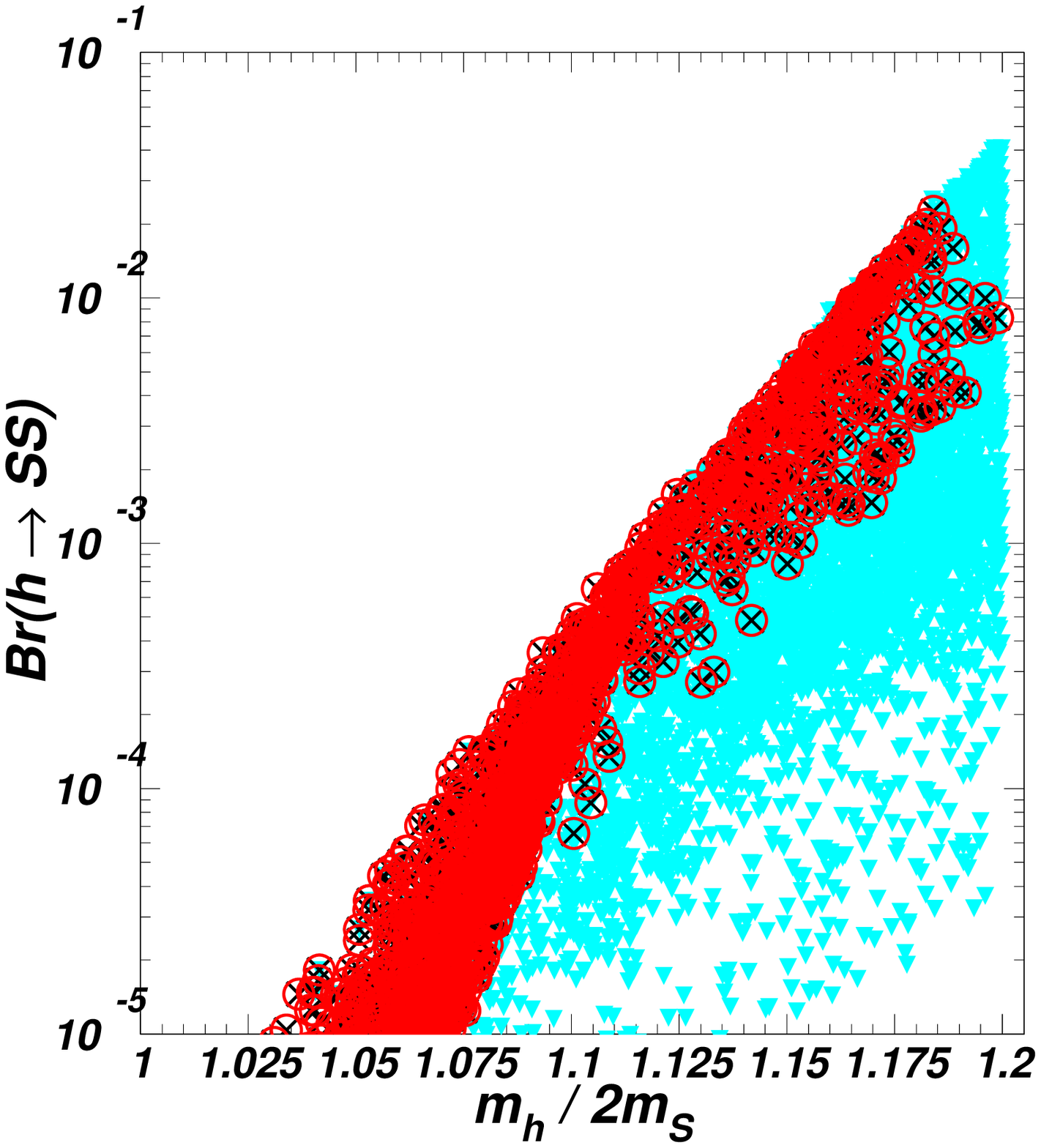,height=7.3cm}
\vspace{-0.5cm} \caption{Same as Fig. \ref{bfigms}, but projected on the planes of 
$Br(h\to SS)$ versus $m_h$ and $Br(h\to SS)$ versus $\frac{m_h}{2m_S}$.}
\label{bfigbr}
\end{figure}

In Fig. \ref{bfigbr}, we project the surviving samples on the
planes of $Br(h\to SS)$ versus $m_h$ and $Br(h\to SS)$ versus $\frac{m_h}{2m_S}$. 
Fig. \ref{bfigbr} shows that in the parameter space allowed by the XENON1T (2017), PandaX-II (2017), and the Fermi-LAT,
$Br(h\to SS)$ is smaller than 3\% for $m_S<$ 60 GeV and smaller than
 0.2\% for 60 GeV $<m_S<$ 120 GeV. The current searches for the DM at 
the LHC do not impose the constraints on the parameter space.

\section{Conclusion}
The type-II 2HDM with a scalar DM provides a WIMP-DM candidate economically.
Recent some studies do not find the parameter space of the DM with a mass below 50 GeV in the model. 
In this paper, we examine the DM with a mass below 50 GeV in the model after 
imposing the constraints from the Higgs searches at the LHC
and the DM experiments.

We first discuss a general scenario in which both two CP-even Higgses ($h$ and $H$) are portals between
the DM and SM sectors, and the CP-odd Higgs ($A$) and $H$ are heavier than 130 GeV. We find that the DM with a mass of $10\sim 50$ GeV
is disfavored by the joint constraints of the 125 GeV Higgs signal data, the relic density, XENON1T (2017), PandaX-II (2017) and
the Fermi-LAT.

Next, we consider a special scenario in which the heavy CP-even Higgs is taken as the 125 GeV Higgs, and
the light CP-even Higgs is the only portal between the DM and SM sectors. The DM mass is slightly
below Higgs resonance, $m_h/2=(1.0\sim 1.2)\times m_S$. We find that the signal data of the 125 GeV Higgs
restrict $\tan\beta$ to be in the range of $1 \sim 1.5$ for $m_h<$ 62 GeV. 
The $gg\to A \to hZ$ channel at the LHC give the lower bounds on $\tan\beta$ for
53 GeV $<m_h<120$ GeV and $m_A=$ 600 GeV. The $b\bar{b}\to h \to \tau^+\tau^-$ channels impose the upper limits on $\tan\beta$, and
$\tan\beta> 10$ is excluded for both $m_h<80$ GeV and $m_h>90$ GeV. For the appropriate values of $\tan\beta$, $\lambda_h$ and $m_h$, 
the DM with a mass of $10\sim 50$ GeV is allowed by the constraints of "pre-LHC", 
the signal data of the 125 GeV Higgs, the searches for the additional Higgs at LHC, the DM relic density,
XENON1T (2017), PandaX-II (2017), and the Fermi-LAT. For example, $\tan\beta$ is restricted to be in
the range of $1.0\sim1.5$ for 10 GeV $<m_s<$ 26 GeV, and $\frac{m_h}{2m_S}>$ 1.125 is excluded for 30 GeV $<m_S<$ 50 GeV.

\section*{Acknowledgment}
We thank Lei Feng for helpful discussions.
 This work is supported by the National Natural Science Foundation
of China under grant No. 11575152, and the Natural Science Foundation of
Shandong province (ZR2017MA004 and ZR2017JL002).


\begin{thebibliography}{99}
\bibitem{1207.4930-1} V. Silveira and A. Zee, \PLB161, 136 (1985).
\bibitem{1412.1105} L. Feng, S. Profumo, L. Ubaldi, \JHEP1503, 045 (2015).
\bibitem{2hisos-6} X.-G. He, J. Tandean, \JHEP1612, 074 (2016).

\bibitem{2hdm} T. D. Lee, \PRD8, 1226 (1973).

\bibitem{2hisos-0} X.-G. He, T. Li, X.-Q. Li, J. Tandean, H.-C. Tsai, \PRD79, 023521 (2009).

\bibitem{2hisos-1} X.-G. He, J. Tandean, \PRD88, 013020 (2013).
\bibitem{2hisos-2} Y. Cai, T. Li, \PRD88, 115004 (2013).
\bibitem{2hisos-3} L. Wang, X.-F. Han, \PLB739, 416-420 (2014).
\bibitem{2hisos-4} A. Drozd, B. Grzadkowski, J. F. Gunion, Y. Jiang, \JHEP1411, 105 (2014).
\bibitem{2hisos-5} A. Drozd, B. Grzadkowski, J. F. Gunion, Y. Jiang, \JCAP1610, 040 (2016).


\bibitem{dmbu} T. Alanne, K. Kainulainen, K. Tuominen, V. Vaskonen, \JCAP1608, 057 (2016). 

\bibitem{1708.06882} L. Wang, R. Shi, X.-F. Han, \PRD96, 115025 (2017). 

\bibitem{1608.00421} N. Chen, Z. Kang, J. Li, \PRD95, 015003 (2017).


\bibitem{2h-poten} R. A. Battye, G. D. Brawn, A. Pilaftsis, \JHEP1108, 020 (2011).

\bibitem{i-1} H. E. Haber, G. L. Kane, T. Sterling, \NPB161, 493
(1979).

\bibitem{ii-2} J. F. Donoghue and L. F. Li, \PRD19, 945 (1979).

\bibitem{sushi} R. V. Harlander, S. Liebler, H. Mantler, \CPC184, 1605 (2013).

\bibitem{bsr580} Heavy Flavor Averaging Group, \EPJC77, 895 (2017);
                 M. Misiak, M. Steinhauser, \EPJC77, 201 (2017).

\bibitem{1701.02678} L. Wang, F. Zhang, X.-F. Han, \PRD95, 115014 (2017). 

\bibitem{1502.07532} M. Gorbahn, J. M. No, V. Sanz, \JHEP1510, 036 (2015).

\bibitem{1604.01406} F. Kling, J. M. No, S. Su, \JHEP1609, 093 (2016). 


\bibitem{2hc-1} D. Eriksson, J. Rathsman, O. St{\aa}l, \CPC181, 189 (2010).

\bibitem{spriso} F. Mahmoudi, \CPC180, 1579-1673 (2009).

\bibitem{deltmq} C. Q. Geng and J. N. Ng, \PRD38, 2857 (1988)
[Erratum-ibid. D 41, 1715 (1990)].

\bibitem{rb1} H. E. Haber, H. E. Logan, \PRD62, 015011 (2000).

\bibitem{rb2} G. Degrassi, P. Slavich, \PRD81, 075001 (2010).


\bibitem{160602266} ATLAS and CMS Collaborations, \JHEP1608, (2016) 045.

\bibitem{47Aad:2014vgg}
 ATLAS Collaboration, G.~Aad {\em et~al.}, ``{Search for neutral
  Higgs bosons of the minimal supersymmetric standard model in pp collisions at
  $\sqrt{s}$ = 8 TeV with the ATLAS detector},'' \JHEP11, 056 (2014).

\bibitem{48CMS:2015mca}
 CMS Collaboration,
``{Search for additional neutral Higgs bosons decaying to a pair of tau leptons
  in $pp$ collisions at $\sqrt{s}$ = 7 and 8 TeV},''
CMS-PAS-HIG-14-029.

\bibitem{82vickey}
 ATLAS Collaboration,
``{Search for Minimal Supersymmetric Standard Model Higgs Bosons $H/A$ in the
  $\tau\tau$ final state in up to 13.3 fb$^{-1}$ of pp collisions at
  $\sqrt{s}$= 13 TeV with the ATLAS Detector},''
ATLAS-CONF-2016-085.

\bibitem{add-hig-16-037}
 CMS Collaboration,
``{Search for a neutral MSSM Higgs Boson decaying into $\tau\tau$ $H/A$ with 12.9 fb$^{-1}$ of data at
  $\sqrt{s}$= 13 TeV},''
CMS-PAS-HIG-16-037.

\bibitem{1709.07242}
ATLAS Collaboration,
``{Search for additional heavy neutral Higgs and gauge bosons in the ditau final state produced in 36 fb$^{-1}$
of pp collisions at $\sqrt{s}$= 13 TeV with the ATLAS detector},''
 \JHEP1801, 055 (2018).


\bibitem{1511.03610}
 CMS Collaboration,
``{Search for a low-mass pseudoscalar Higgs boson produced
in association with a $b\bar{b}$ pair in pp collisions at $\sqrt{s}$ = 8 TeV},''
 \PLB758, 296-320 (2016).


\bibitem{CMS-HIG-15-009}
 CMS Collaboration,
``{Search for a light pseudoscalar Higgs boson produced in association with bottom
quarks in pp collisions at $\sqrt{s}$ = 8 TeV},''
CMS-HIG-15-009.


\bibitem{80lenzi}
 ATLAS Collaboration,
``{Search for scalar diphoton resonances with 15.4~fb$^{-1}$ of data collected
  at $\sqrt{s}$=13 TeV in 2015 and 2016 with the ATLAS detector},''
ATLAS-CONF-2016-059.

\bibitem{81rovelli}
 CMS Collaboration,
``{Search for resonant production of high mass photon pairs using
  $12.9\,\mathrm{fb^{-1}}$ of proton-proton collisions at $\sqrt{s} =
  13~\mathrm{TeV}$ and combined interpretation of searches at 8 and 13 TeV},''
CMS-PAS-EXO-16-027.

\bibitem{HIG-17-013-pas}
 CMS Collaboration,
``{Search for new resonances in the diphoton final state in the mass range between
70 and 110 GeV in pp collisions at $\sqrt{s}$ = 8 and 13 TeV},''
CMS-PAS-HIG-17-013.


\bibitem{55Aad:2015agg}
 ATLAS Collaboration, G.~Aad {\em et~al.}, ``{Search for a high-mass
  Higgs boson decaying to a $W$ boson pair in $pp$ collisions at $\sqrt{s} = 8$
  TeV with the ATLAS detector},'' \JHEP01, (2016) 032.


\bibitem{77atlasww13}
 ATLAS collaboration,
``{Search for a high-mass Higgs boson decaying to a pair of W bosons in pp
  collisions at $\sqrt{s}=13$ TeV with the ATLAS detector},'' ATLAS-CONF-2016-074.

\bibitem{78atlasww13lvqq}
 ATLAS Collaboration, ``Search for diboson resonance production
in the $\ell\nu qq$ final state using p p collisions at $\sqrt{s}$ = 13 TeV
with the ATLAS detector at the LHC,''
ATLAS-CONF-2016-062.


\bibitem{1710.07235}
 ATLAS Collaboration, 
``Search for WW/WZ resonance production in $\ell\nu qq$ final states in pp collisions 
at $\sqrt{s}$ = 13 TeV with the ATLAS detector,''
arXiv:1710.07235.

\bibitem{1710.01123}
ATLAS Collaboration, 
``Search for heavy resonances decaying into WW
in the $e\nu\mu\nu$ final state in pp collisions $\sqrt{s}$ = 13 TeV with the ATLAS detector,''
\EPJC78, 24 (2018).



\bibitem{57Aad:2015kna}
 ATLAS Collaboration, G.~Aad {\em et~al.}, ``{Search for an
  additional, heavy Higgs boson in the $H\rightarrow ZZ$ decay channel at
  $\sqrt{s} = 8\;\text{ TeV }$ in $pp$ collision data with the ATLAS
  detector},'' \EPJC76, 45 (2016).

\bibitem{74koeneke4}
 ATLAS Collaboration,
``{Search for new phenomena in the $Z(\rightarrow\ell\ell) +
  E_{\mathrm{T}}^{\mathrm{miss}}$ final state at $\sqrt{s}$ = 13 TeV with thee
  ATLAS detector},''
ATLAS-CONF-2016-056.

\bibitem{75koeneke5}
ATLAS Collaboration,
``{Searches for heavy ZZ and ZW resonances in the $\ell\ell qq$ and vvqq final states in
  pp collisions at $\sqrt{s} = 13$ TeV with the ATLAS detector},''
ATLAS-CONF-2016-082.

\bibitem{76koeneke3}
 ATLAS Collaboration,
``{Study of the Higgs boson properties and search for high-mass scalar
  resonances in the $H \rightarrow ZZ^* \rightarrow 4\ell$ decay channel at
  $\sqrt{s}$ = 13 TeV with the ATLAS detector},''
ATLAS-CONF-2016-079.


\bibitem{1712.06386}
 ATLAS Collaboration,
``{Search for heavy ZZ resonances in the $\ell^+\ell^-\ell^+\ell^-$ and $\ell^+\ell^-\nu\nu$ final states 
using proton proton collisions at $\sqrt{s}$ = 13 TeV with the ATLAS detector},''
arXiv:1712.06386.

\bibitem{1708.09638}
ATLAS Collaboration,
``{Searches for heavy ZZ and ZW resonances in the $\ell\ell qq$ and $\nu\nu qq$ final states in pp collisions 
at $\sqrt{s}$ = 13 TeV with the ATLAS detector},''
arXiv:1708.09638.




\bibitem{64Khachatryan:2016sey}
 CMS Collaboration, V.~Khachatryan {\em et~al.}, ``{Search for two
  Higgs bosons in final states containing two photons and two bottom quarks},''
\PRD94, 052012 (2016).

\bibitem{65Khachatryan:2015yea}
 CMS Collaboration, V.~Khachatryan {\em et~al.}, ``{Search for
  resonant pair production of Higgs bosons decaying to two bottom
  quark--antiquark pairs in proton--proton collisions at 8 TeV},''
 \PLB749, 560-582 (2015).

\bibitem{66Khachatryan:2015tha}
 CMS Collaboration, V.~Khachatryan {\em et~al.}, ``{Searches for a
  heavy scalar boson H decaying to a pair of 125 GeV Higgs bosons hh or for a
  heavy pseudoscalar boson A decaying to Zh, in the final states with $h \to
  \tau \tau$},'' \PLB755, 217-244 (2016).

\bibitem{add-CONF-2016-004}
 ATLAS Collaboration, ``{Search for Higgs boson pair production in the
$b\bar{b}\gamma\gamma$ final state using pp collision data at $\sqrt{s}=13$ TeV with
the ATLAS detector},''
ATLAS-CONF-2016-004.

\bibitem{add-hig-16-032}
 CMS Collaboration, V.~Khachatryan {\em et~al.}, ``{Search for
H($b\bar{b}$)H($\gamma\gamma$) decays at $\sqrt{s}=13$ TeV},''
CMS-PAS-HIG-16-032.

\bibitem{84varol}
 ATLAS Collaboration,
``{Search for pair production of Higgs bosons in the $b\bar{b}b\bar{b}$ final
  state using proton$-$proton collisions at $\sqrt{s} = 13$ TeV with the ATLAS
  detector},'' ATLAS-CONF-2016-049.

\bibitem{85CMS:2016knm}
 CMS Collaboration,
``{Search for resonant Higgs boson pair production in the
 $\mathrm{b\overline{b}}\tau^+\tau^-$ final state using 2016 data},''
CMS-PAS-HIG-16-029.

\bibitem{1710.04960}
 CMS Collaboration,
``{Search for a massive resonance decaying to a pair of Higgs bosons in the four b quark final state in proton-proton collisions
 at $\sqrt{s} = 13$},''
arXiv:1710.04960.

\bibitem{1707.02909}
 CMS Collaboration,
``{Search for Higgs boson pair production in events with two bottom quarks and two tau leptons in proton-proton collisions 
at $\sqrt{s} = 13$},''
arXiv:1707.02909.



\bibitem{67Khachatryan:2015lba}
 CMS Collaboration, V.~Khachatryan {\em et~al.}, ``{Search for a
  pseudoscalar boson decaying into a $Z$ boson and the 125 GeV Higgs boson in
  $\ell^+\ell^-b\overline{b}$ final states},''
  \PLB748, 221-243 (2015).

\bibitem{68Aad:2015wra}
 ATLAS Collaboration, G.~Aad {\em et~al.}, ``{Search for a CP-odd
  Higgs boson decaying to Zh in pp collisions at $\sqrt{s} = 8$ TeV with the
  ATLAS detector},'' \PLB744, 163-183 (2015).

\bibitem{69AZhatlas13}
 ATLAS Collaboration,
``{Search for a CP-odd Higgs boson decaying to Zh in pp collisions at $\sqrt{s}$ = 13
  TeV with the ATLAS detector},'' ATLAS-CONF-2016-015.

\bibitem{1712.06518}
 ATLAS Collaboration,
``{Search for heavy resonances decaying into a W or Z boson and a Higgs boson in final states with leptons and b-jets 
in 36 $fb^{-1}$ of $\sqrt{s}$ = 13 pp collisions with the ATLAS detector},''
 arXiv:1712.06518.

\bibitem{1505.01609}
 ATLAS Collaboration,
``{Search for Higgs bosons decaying to aa in the $\mu\mu\tau\tau$ final state in pp collisions 
at $\sqrt{s}$= 8 TeV with the ATLAS experiment},'' \PRD92, 052002 (2015).

\bibitem{1701.02032}
 CMS Collaboration,
``{Search for light bosons in decays of the 125 GeV Higgs boson in proton-proton collisions at $\sqrt{s}$= 8 TeV},''
\JHEP1710, 076 (2017).

\bibitem{160302991}
CMS Collaboration, V.~Khachatryan {\em et~al.},
``{Search for neutral resonances decaying into a Z boson and
a pair of b jets or $\tau$ leptons},'' 
\PLB759, 369-394 (2016).


\bibitem{hb1} P. Bechtle, O. Brein, S. Heinemeyer, G. Weiglein, K. E.
Williams, \CPC181, 138-167 (2010).

\bibitem{hb2} P. Bechtle, O. Brein, S. Heinemeyer, O. St{\aa}l, T. Stefaniak, G. Weiglein, K. E. Williams,
\EPJC74, 2693 (2014).

\bibitem{1608.02573} R. K. Barman, B. Bhattacherjee, A. Choudhury, D. Chowdhury, 
J. Lahiri, S. Ray, arXiv:1608.02573.


\bibitem{mhp500} S. Moretti, arXiv:1612.02063.

\bibitem{1706.07414} A. Arbey, F. Mahmoudi, O. Stal, T. Stefaniak, arXiv:1706.07414.


\bibitem{micomega} G. Belanger, F. Boudjema, A. Pukhov, A. Semenov,
\CPC185, 960-985 (2014).

\bibitem{feyrule} A. Alloul et al., \CPC185, 2250 (2014).

\bibitem{sigis} G. Jungman, M. Kamionkowski, K. Griest, \PR267, 195
(1996); M. A. Shifman, A. I. Vainshtein, V. I. Zakharov, \PLB78,
443 (1978).


\bibitem{planck}
Planck Collaboration, \ASA27, 594 (2016).

\bibitem{pandax} PandaX Collaboration, \PRL119, 181302 (2017).

\bibitem{xenon} E. Aprile et al. [XENON Collaboration], \PRL119, 181301 (2017).

\bibitem{fermi} Fermi-LAT Collaboration, \PRL115, 231301 (2015).


\end{thebibliography}
\end{document}